\documentclass[superscriptaddress,groupedaddress,nofootnoteinbib,11pt]{article}  
\usepackage{graphicx}  
\usepackage{dcolumn}   
\usepackage{bm}        
\usepackage{amssymb}   
\usepackage{feynmf}	
\usepackage{jcappub}

\def\be{\begin{equation}}
\def\ee{\end{equation}}
\def\ba{\begin{eqnarray}}
\def\ea{\end{eqnarray}}
\def\beq{\begin{eqnarray}}
\def\eeq{\end{eqnarray}}

\def\mpl{M_{\rm Pl}}

\def\d{\mathrm{d}}
\def\p{{\cal P}}

\def\K{{\cal K}}

\def\L*{{\cal L}_*}
\def\L{\mathcal{L}}
\def\({\left(}
\def\){\right)}
\def\ie{{\it i.e. }}

\def\nn{\nonumber}
\def\p{\partial}
\def\mn{_{\mu \nu}}
\def\stu{St\"uckelberg }
\def\p{\partial}
\def\mupn{^\mu_{\ \nu}}
\def\<{\langle}
\def\>{\rangle}
\def\Ein{\hat{\mathcal{E}}}
\def\hel{{\rm helicity}}
\def\st{{\rm Stueckelberg}}

\begin{document}

\title{Helicity Decomposition of Ghost-free \\ Massive Gravity}

\author[1,2]{Claudia de Rham,}
\author[3]{Gregory Gabadadze}
\author[2]{and Andrew J. Tolley}
\affiliation[1]{D\'epartment de Physique  Th\'eorique and Center for Astroparticle Physics, Universit\'e
de  Gen\`eve, 24 Quai E. Ansermet, CH-1211  Gen\`eve}
\affiliation[2]{Department of Physics, Case Western Reserve University, 10900 Euclid Ave, Cleveland, OH 44106, USA}
\affiliation[3]{Center for Cosmology and Particle Physics,
Department of Physics, New York University,
NY, 10003, USA}


\abstract{
 We perform a helicity decomposition in the full Lagrangian of the class of Massive Gravity
theories previously proven to be free of the sixth (ghost) degree of freedom via a Hamiltonian
analysis. We demonstrate, both with and without the use of nonlinear field redefinitions, that the scale at which the first interactions of the helicity-zero mode come in is $\Lambda_3=(\mpl m^2)^{1/3}$, and that this is the same scale at which helicity-zero perturbation theory breaks down. We show that the number of propagating  helicity modes remains five in the full nonlinear theory with sources. We clarify recent misconceptions in the literature advocating the existence of either a ghost or a breakdown of perturbation theory at the significantly lower energy scales, $\Lambda_5=(\mpl m^4)^{1/5}$ or $\Lambda_4=(\mpl m^3)^{1/4}$,
which arose because relevant terms in those calculations were overlooked.
As an interesting byproduct of our analysis, we show that it is possible to derive the \stu
formalism  from the helicity decomposition, without ever invoking diffeomorphism invariance, just
from a simple requirement that the kinetic terms of the helicity-two, -one and -zero modes are diagonalized.}

\maketitle


\section{Introduction}

General Relativity (GR), with or without a cosmological constant, is the unique theory of a single massless spin-2 field in four dimensions. Consistent with the requirements for a massless spin-2 representation of the Poincar\'e group, it only excites the helicity-2 modes of the spin-2 field, while the helicity-1 and 0 modes are simply absent. This statement is true to all orders in interactions, and  is made manifest in the standard
formulation of GR by the existence of a local symmetry: nonlinear diffeomorphism invariance. This symmetry ensures that any potential additional helicity-1 and helicity-0 modes are pure gauge degrees of freedom.

If the spin-2 field acquires a mass, then five polarizations (two helicity-2, two helicity-1 and one helicity-0) are necessarily excited in a theory which preserves Lorentz invariance. At the same time, diffeomorphism invariance is broken by the mass term. This means the interactions in a massive spin-2 theory, both with itself and with matter, are less constrained than in a massless one. As a consequence, it has frequently been argued that a sixth ghostly mode would arise in any attempt to define an interacting theory of a massive spin-2 field. The sixth mode shows up as higher derivative interactions for the required five polarizations. For a Fierz-Pauli theory, the mass term is defined such that this sixth mode is absent when considering quadratic fluctuations around flat space-time, \cite{Fierz:1939ix}. However the same does not necessarily remain true around an arbitrary background, or at higher orders in perturbations, and until recently it was believed that massive gravity would inexorably excite a sixth
(ghost) mode non-linearly. This was the state of affairs until a specific model of massive gravity was proposed in \cite{deRham:2010kj} which generalizes the Fierz-Pauli mass to all orders, \cite{deRham:2010ik}. That model was shown to be free of the ghost degree of freedom, in full generality in the ADM formalism \cite{Hassan:2011hr} following earlier arguments in \cite{deRham:2010kj}, and in the \stu language both in the decoupling limit \cite{deRham:2010kj,deRham:2010ik} and beyond \cite{deRham:2011rn}.

Massive gravity was recently reanalyzed using a standard helicity decomposition of the massive spin-2 field \cite{Folkerts:2011ev}, and despite the above results it was suggested that a ghost, or failure of perturbation theory for the spin-2 field occurred at a surprisingly low scale in which the theory was already known to be effectively free via the \stu analysis and ghost free via the ADM analysis. In this paper, we reconsider the helicity argument, and show that when all the terms in the Lagrangian are properly accounted for at a given energy scale (including the ones left out
in \cite{Folkerts:2011ev}), the massive gravity theory proposed in \cite{deRham:2010kj} is free from both ghost and low strong-coupling scale problems. We are able to reconfirm all previously known results directly in the helicity variables without performing any field redefinitions and clarify how they are preserved even when considering the coupling to matter. We clarify the connection between the \stu and helicity decompositions and explain why the field redefinition between the two is well-defined and remains under perturbative control below the scale $\Lambda_3=(\mpl m^2)^{1/3}$. We also show explicitly how the \stu decomposition can be derived from the helicity one by the need to diagonalize the kinetic terms of the different helicity modes around arbitrary backgrounds. These conclusions are in complete consistency with the results previously obtained in unitary gauge which showed that the theory was ghost free, \cite{deRham:2010kj,Hassan:2011hr}.

Before focusing on this helicity argument, it is worth putting this paper in context and
summarizing the ups and downs in the  development of massive gravity. A recent review of massive gravity can also be found in 
\cite{Hinterbichler:2011tt}.
Readers familiar with those works
may skip to section \ref{skiphere}. We recall here the different ways a generic model
of massive gravity has been argued to contain ghosts, and emphasize how the model presented in
\cite{deRham:2010kj} evades all these arguments.
For simplicity, the energy scale $\Lambda_n$ at which a given interaction occurs is
expressed as $\Lambda_n=(\mpl m^{n-1})^{1/n}$.\\

{\bf A. Absence of the Ghost in Unitary gauge}:\\

\noindent Since a massive spin-2 field has five polarizations, while a massless one only carries two, the linear 
Fierz-Pauli theory has a discontinuity in the limit of vanishing mass as was first shown  by van Dam,
Veltman, and Zakharov (vDVZ discontinuity), \cite{vDVZ}. This discontinuity was argued
to rule out massive gravity on observational grounds. However, soon after Vainshtein
showed that  theories of massive gravity
could  avoid the vDVZ discontinuity  problem due to the existence of nonlinear interactions, \cite{Arkady}. The question
then was whether these nonlinear interactions could consistently be incorporated into
the Fierz-Pauli  theory. It was then realized that massive gravity could potentially suffer from ghost-like pathologies
due to a possible sixth degree of freedom emerging
at the non-linear level, as first shown by  Boulware and Deser, who discovered
that the lapse $N^0$ always enters non-linearly for any
mass term which is a function of the Fierz-Pauli combination $f(h\mn^2-h^2)$
or of the metric determinant, \cite{Boulware:1973my}.

Whilst the results of Boulware and Deser are correct, they did not exhaust all possible interactions for massive gravity theories. Most importantly, they do not account for a subtlety  which allows the theory to maintain a Hamiltonian constraint even when the lapse enters non-linearly. More precisely, in GR, the lapse $N^0$ and the shift $N^i$ both enter linearly in the Hamiltonian, and it is therefore transparent that this theory has a Hamiltonian constraint. In massive gravity, the lapse always enters non-linearly, so differentiating with respect to the lapse does not directly enforce a constraint, but rather an equation through which the lapse itself could in principle be determined. However, to be able to express all the shifts and the lapse through these equations of motion, the system of corresponding equations should be invertible, or equivalently, the determinant of the Hessian $\mathcal{H}_{ab}=\p^2\L_m/ \p N^a \p N^b$  (where $N^a$ collectively denotes the shift and lapse)
should not vanish.  In the model of massive gravity proposed in \cite{deRham:2010kj}, (called ``Ghost-free Massive Gravity" in what follows) it has been shown explicitly that this condition does not hold and the  determinant of the Hessian is zero (equivalently it has been shown that the Hamiltonian is linear in the lapse after integration over the shift),
\cite{deRham:2010kj,Hassan:2011hr}. As a consequence, not all the equations of motion are independent, but instead one of them
gives rise to a constraint. This constraint is what projects out the potential sixth degree of freedom,
the so-called Boulware-Deser (BD) ghost. This suggests that one should look for a reformulation of the theory
in which a non-linearly redefined lapse enforces a constraint from the outset \cite{deRham:2010kj,Hassan:2011hr}.

The full unitary gauge ADM analysis performed \cite{Hassan:2011hr} confirms the absence of ghosts in the class of models proposed in  \cite{deRham:2010kj}. However, in this approach the energy scale of the interactions of the different helicity modes is not transparent. This is the reason why alternative approaches that introduce auxiliary fields compensated by additional gauge symmetries have been utilized to analyze the interactions, principally to determine at which scale the helicity-zero mode becomes strongly coupled. The two main approaches are the \stu approach, which introduces auxiliary fields compensated by full nonlinear diffeomorphisms, and the helicity approach which introduces auxiliary fields compensated by only linear diffeomorphisms and an additional $U(1)$ symmetry, described below. We should stress however that since the unitary gauge formulation is ghost free, any apparent ghost that appears via the introduction of auxiliary fields cannot actually be present. \\

{\bf B. Absence of the Ghost in the \stu language}:\\

\noindent In 2002, Arkani-Hamed, Georgi and Schwartz (AGS), suggested a complementarily approach to study massive gravity as an effective field theory and introduced four \stu fields $\phi^a$ to make the counting of degrees of freedom transparent, \cite{ArkaniHamed:2002sp}, (see also \cite{Dubovsky}). In this approach, the covariance of the theory is also explicit, but the power  of this framework is in its ability to identify the relevant degrees of freedom present in the theory, even when they arise at different scales. This is possible
by taking a specific limit of the theory  in such a way that the different degrees of freedom decouple,
hence called the ``decoupling limit".
\begin{enumerate}
\renewcommand{\labelenumi}{(\alph{enumi})}
\item{\bf In the decoupling limit}: In this limit, AGS showed that the ghost could in principle be pushed beyond the scale $\Lambda_5$, however shortly after, it was argued that the ghost would inexorably reappear at the scale $\Lambda_3$, \cite{Creminelli:2005qk}. This was later shown not to be the most general answer, and there exists a class of theories  (corresponding to Ghost-free Massive Gravity) for which
     not only does the decoupling limit occur at the scale $\Lambda_3$, but no ghost is present at that scale, \cite{deRham:2010ik,deRham:2010kj}, the decoupling limit if therefore completely healthy. 
\item{\bf Beyond the decoupling limit:} It the full theory the absence of the BD ghost was
shown up to and including the quartic order in nonlinearities in \cite{deRham:2010kj},
and to all orders in \cite{Hassan:2011hr}. However, a
analysis of the theory at energy scales beyond $\Lambda_3$ 
indicated the appearance of $(\dot \phi^{0})^2$ which have been interpreted as the revival
of the BD ghost at larger scales, \cite{slavapaper1,slavapaper2}, (notice that these are not the only terms appearing at the that scale and that order).

Whilst the presence of such a term is manifest, the conclusions of
\cite{slavapaper1,slavapaper2} on the existence of the BD ghost are not correct as these works
do not account for the existence of a constraint, which removes the ghost, \cite{deRham:2011rn}.
The fact that  $\dot \phi^0$ comes in non-linearly in the action does indeed imply that the equation of motion
with respect to $\phi^0$ is dynamical as it involves some $\ddot \phi^0$, however this does not
yet imply that all the four \stu fields are independent dynamical degrees of freedom. This would
only be the case if the determinant of the Hessian
$\mathcal{H}_{ab}=\p^2\L_m/ \p \dot \phi^a \p \dot \phi^b$ was nonzero. However it has been
shown explicitly that for the specific mass term considered in Ghost-free Massive Gravity,
this condition did not hold and therefore not all four \stu fields are independent even
beyond the decoupling limit, \cite{deRham:2011rn}.
This theory therefore propagates 3 degrees of freedom out of four $\phi^a$   \stu fields on top of
the 2 usual tensor polarizations, leading to 5 degrees of freedom; this  is the correct counting
in the absence of the BD ghost.

The existence of this constraint manifests itself in various ways in classical solutions discussed
in \cite{deRham:2010tw,Koyama,datogiga,cosmology}.
\end{enumerate}\vspace{10pt}

{\bf C. (Non-)existence of the Ghost in the Helicity language}: \\

\noindent Finally, despite the proof for the absence of any ghost in the model \cite{deRham:2010kj} both in unitary gauge in the ADM language and the \stu approach (first in the decoupling limit and then later beyond it), it has recently been argued, that the ghost actually manifests itself in the helicity decomposition, at the scale $\Lambda_4$ already at cubic order in perturbations, and then at the even lower scale $\Lambda_5$ at quartic order in perturbations, \cite{Folkerts:2011ev}. Whilst this result is in clear contradiction with the results present in the literature, we here analyze this argument in more detail and show once again the absence of ghost in the
helicity language when all terms in the Lagrangian are properly accounted for (including terms previously left out in
\cite{Folkerts:2011ev}). \\

\subsection{Relationship between the Helicity and \stu Decompositions}

\label{skiphere}

Since this paper is concerned with the helicity decomposition, let us clarify the relationship between this and the now familiar \stu decomposition. Both decompositions rely on splitting the massive spin-2 field up into a would-be helicity-2 $\hat{h}_{\mu\nu}$, helicity-1 $A_{\mu}$ and helicity-0 modes $\pi$.
In other words, both decompositions express the spin-2 metric perturbation in the form
\ba
h\mn=\mpl (g\mn-\eta\mn) = \hat{h}\mn + \frac{1}{2m}\partial_{(\mu} A_{\nu )} + \frac{1}{3m^2} \partial_{\mu}\partial_{\nu} \pi + D\mn\,,
\ea
where the final piece $D\mn$ is chosen to diagonalize the kinetic term of the different helicities to avoid kinetic mixing.
In both decompositions the helicity-1 and helicity-0 terms are identified in the same way
\ba
\pi^\hel = \pi^\st \\
A_{\mu}^\hel =A_{\mu}^\st\,.
\ea
The difference lies entirely in the final term $D\mn$, or equivalently in the identification of the helicity-2 mode. In the {\bf helicity decomposition} this is chosen to diagonalize only the free theory (Fierz-Pauli) kinetic term defined around Minkowski spacetime
\ba
D^\hel\mn = \frac 16  \pi  \eta\mn.
\ea
By contrast, in the \stu decomposition, one chooses $D\mn$ to diagonalize the kinetic mixing term between the different helicities for all terms that arise at energy scales below $\Lambda_3$ {\bf around an arbitrary background}. Direct calculation (performed below) shows that this gives\footnote{In most presentations of the \stu approach, the diagonalization term $D_{\mu \nu}=\frac{1}{6}\pi \eta_{\mu \nu}$ is not included until the very end of the calculation. For ease of comparison with the helicity calculation we include it from the outset. None of the conclusions of standard \stu presentations are affected by this.}
\ba
&&D^\st \mn = \frac 16  \pi\eta\mn + \mpl \(\Psi_{\mu\alpha} \Psi_{\nu}{}^{\alpha}-\eta_{\mu\nu}- \frac{1}{2m}\partial_{(\mu} A_{\nu )} - \frac{1}{3m^2} \partial_{\mu}\partial_{\nu} \pi \) \\
&=&  \frac{1}{6} \eta_{\mu\nu} \pi + \frac{1}{36 \Lambda_5^5} \partial_{\mu} \partial_{\alpha} \pi \partial_{\nu} \partial^{\alpha} \pi + \frac{1}{12\Lambda_4^4} \partial_{(\mu}A^{\alpha} \partial_{\nu)} \partial_{\alpha} \pi    + \frac{1}{4 \Lambda_3^3} \partial_{\mu}A_{\alpha} \partial_{\nu} A^{\alpha} \nn\,,
\ea
where $\Psi\mn = \eta\mn+\frac{1}{2\mpl  m}\partial_{\mu} A_\nu + \frac{1}{6 \mpl m^2} \p_\mu \p_\nu \pi $.
The \stu decomposition does not completely diagonalize the kinetic terms, it leaves a mixing that arises at the scale $\Lambda_3$ . One could go further and diagonalize at these scales, however it is known that for the generic two parameter allowed massive gravity models the diagonalization needed is nonlocal at these scales (although it is local for special cases). \\

In the \stu decomposition, the diagonalization term $D\mn$ is defined nonlinearly in the helicity fields. More precisely $(D\mn^\st  -D^\hel\mn) $ is quadratic in the fields.
This does not mean that performing this field redefinition implies perturbation theory for a spin-2 field is breaking down. {\bf A direct application of perturbation theory in the helicity variables without the use of this field redefinition shows that for energy scales below $\Lambda_3$ perturbation theory always converges}.
Since there is no fundamental need to diagonalize the kinetic terms (it simply simplifies calculations) we will find that the helicity and \stu decompositions agree identically on the energy scale at which interactions come in. In particularly, both decompositions will agree that the first interactions that arise are at the scale $\Lambda_3$ and that the helicity-0 mode becomes strongly coupled precisely at this scale. This is all much easier to see in the \stu language because of the choice to diagonalize all kinetic mixings below the scale $\Lambda_3$ around an arbitrary background. Nevertheless, with some effort (see below) all the results are confirmed by the helicity decomposition. \\

In summary, it is the helicity-2 mode (and not the helicity-0) which is defined differently in each language
\ba
h\mn^\st  = h\mn^\hel  +\frac{1}{36 \Lambda_5^5} \p_\mu \p_\alpha \pi \p_\nu \p^\alpha \pi-\frac{1}{12\Lambda_4^4}  \p_{(\mu}A^{\alpha} \p_{\nu)} \p_{\alpha} \pi  + \frac{1}{4 \Lambda_3^3} \p_\mu A_\alpha \p_\nu A^\alpha.
\ea
Crucially this field redefinition is trivially invertible to all orders and includes only a finite number of terms. Furthermore since the second term is nonlinear, this redefinition does not change the asymptotic states in scattering amplitudes, and thus both $h\mn^\st $ and $ h\mn^\hel $ deserve the right to be called the helicity-2 field. The fact that the redefinition includes terms at the scales $\Lambda_5$ and $\Lambda_4$ may cause one to suspect that it disguises interactions at these scales or that perturbation theory breaks down at these scales. We shall find by direct calculation in the helicity language that neither of these two fears is upheld and that perturbation theory of the entire spin-2 field only breaks down at the scale $\Lambda_3$. Thus the apparent hierarchy of new interactions at the scales $\Lambda_5$, $\Lambda_4$ and $\Lambda_3$ is a fake one.
\\

The rest of this paper is organized as follows: In section \ref{sec:ToyModel} we present a scalar field toy-model that summarizes the essence of the argument.  We then move onto discussing the allowed two parameter family of ghost-free models of interacting spin-2 fields, i.e. massive gravity, in section \ref{sec:model} before proving the absence of ghost at the scale $\Lambda_5$ in section \ref{sec:L5}, then at the scale $\Lambda_4$ in section \ref{sec:L4} and finally all the way up to the scale $\Lambda_3$ and beyond in section \ref{sec:L3}. We emphasize how the \stu decomposition emerges from this framework without ever evoking diffeomorphism invariance by the simple requirement of diagonalizing the kinetic term of the different helicity modes. Throughout we emphasize why these results are left unchanged by the coupling to matter.

\section{A Two Scalar Field Toy-Model}
\label{sec:ToyModel}

Before jumping into the subtleties of massive gravity, let us consider a representative scalar field model which will capture the essential points. The essential problem we are faced with is that when massive gravity is analyzed in the helicity decomposition, it appears to generate higher derivative interactions at scales which are significantly lower than when the same analysis is performed in the \stu decomposition. In \cite{Folkerts:2011ev} this fact was used to argue either for the presence of a ghost at these low scales or for the breakdown of spin-2 perturbation theory. Since the relationship between the helicity and \stu decompositions is just a field redefinition (described above), the two methods could only disagree if the field redefinition were ill-defined\footnote{It is worth emphasizing however that since both descriptions are equivalent in Unitary gauge, they {\bf cannot} disagree on the existence of a ghost, even if the field redefinition were ill-defined, they could only disagree on the scale of strong coupling.}. The key to seeing that the field redefinition is well-defined is that it is not a redefinition of a single field in terms of another single field, but is rather a mixing of two fields which arise when the kinetic term of one field is diagonalized. As such we will be able to see both in the helicity and \stu decompositions that there are the same number of degrees of freedom even in the presence of a source. This guarantees the absence of ghosts. Furthermore we will see that perturbation theory remains under control since the perturbative expansion always converges below the scale $\Lambda_3$. This property is again tied to the fact that the field redefinition between the helicity and \stu decompositions is one that mixes two fields in a manner characteristic of a kinetic term diagonalization.

The field redefinition that relates the helicity and \stu decompositions allows us to write a theory which is already perturbatively unitary in a more manifest way, and this can be performed to all orders in the fields expansion in a way that preserves the consistency and validity of perturbation theory. In other words whilst the field redefinition from the helicity to the \stu decomposition is helpful, it is not fundamentally necessary. We can, and will below, understand all the physics directly in terms of the helicity variables and without the use of field redefinitions see that perturbation theory does not break down until the scale $\Lambda_3$ in complete agreement with the \stu decomposition. \\

We summarize this procedure in a simple two scalar field model. In this example, as we have explained, it is essential to maintain a coupling between two different fields, (as is naturally the case in massive gravity). We consider a field $\phi$ (which mimics the helicity-0 mode in massive gravity) and another field $\psi$ (that mimics the helicity-2).
As a simple but representative example, one can consider the (assumed to be exact) tree-level Lagrangian
\ba
\label{toy}
\mathcal{L}=\frac 12 \phi\Box \phi+\frac 12 \psi\Box \psi + \frac{\Box \psi (\p_\alpha \p_\beta \phi)^2}{\Lambda^5}  + \frac{(\p_\alpha \p_\beta \phi)^2\Box (\p_\mu\p_\nu \phi)^2}{2\Lambda^{10}}\,.
\ea
By integrating w.r.t. the field $\psi$, one gets $\psi = - (\partial_\alpha\partial_\beta \phi)^2/\Lambda^5
+\psi_0$, where $\square \psi_0=0$. Substituting this back into the Lagrangian, the latter
reduces  (up to  a total derivative)  to a free theory for $\phi$. Hence, two initial data
is needed to determine $\phi$,  and another two to determine  $\psi$,  leaving no room for
any extra ghostly degree of freedom.

On the other hand, looking at (\ref {toy}) in its entirety,  it does not appear  to be
obviously safe:
{\it A priori} the conjugate momentum to $\psi$ does not seem well defined, however this can be resolved
using a standard Ostrogradsky argument as is performed in the appendix \ref{appendix}.
The fact that this theory involves higher derivative interactions seems to imply, at first glance,
the breakdown of perturbative unitarity at the scale $\Lambda$. Usually this is indicative of the
presence of ghosts at that scale since in most theories the breakdown of perturbative unitarity
also signals the breakdown of genuine unitarity.

However as we shall see, here unitarity is perfectly intact. To see this let us calculate some scattering amplitudes. Whilst the 3-point function $\langle \psi \phi \phi\rangle$ vanishes on-shell from momentum conservation,  the 4-point function $\langle \phi^4\rangle$ on the other hand receives a contribution from the vertex $\p^{10}\phi^4 / \Lambda^{10}$ which taken alone would indeed break perturbative unitarity at the scale $\Lambda$. However, in addition,  there exist three other diagrams that contribute to this amplitude, which propagate a virtual $\psi$, through a decay channel of the form $ \phi \phi \to \psi \to \phi \phi$. Such a diagram involves two vertexes, each of the form $\p^6 \psi \phi^2 /\Lambda^5$, which by themselves also break unitarity. However, one can easily see that the sum of all four diagrams vanishes. \vspace{20pt}
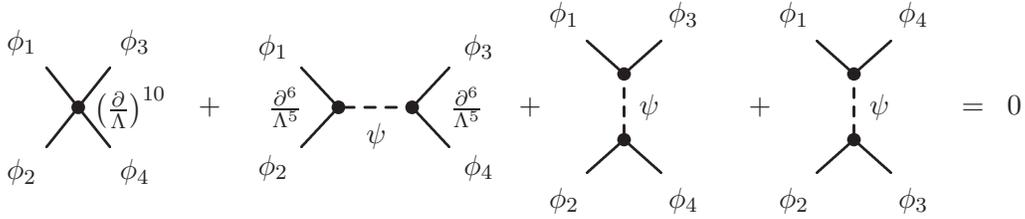
\begin{figure}[!htb]
\begin{center}
\label{phi4}
\begin{fmffile}{ToyModel}
\parbox{20mm}{\begin{fmfgraph*}(30,30)
	            \fmfleft{i1,i2}
	            \fmfright{o1,o2}
                \fmflabel{$\phi_2$}{i1} 
                \fmflabel{$\phi_1$}{i2} 
                \fmflabel{$\phi_4$}{o1} 
                \fmflabel{$\phi_3$}{o2} 
                \fmf{plain}{i1,v1}
                \fmf{plain}{i2,v1}
                \fmf{plain}{v1,o1}
                \fmf{plain}{v1,o2}
                \fmfdot{v1}
                \fmflabel{$\(\frac{\p}{\Lambda}\)^{10} $}{v1}
\end{fmfgraph*}} +\hspace{20pt}
\parbox{30mm}{\begin{fmfgraph*}(70,30)
	            \fmfleft{i1,i2}
	            \fmfright{o1,o2}
                \fmflabel{$\phi_2$}{i1} 
                \fmflabel{$\phi_1$}{i2} 
                \fmflabel{$\phi_4$}{o1} 
                \fmflabel{$\phi_3$}{o2} 
	            \fmf{dashes,label=$\psi$}{v1,v2}
                \fmf{plain}{i1,v1}
                \fmf{plain}{i2,v1}
                \fmf{plain}{v2,o1}
                \fmf{plain}{v2,o2}
                \fmfdot{v1}
                \fmfdot{v2}
                \fmflabel{$\frac{\p^6}{\Lambda^5} \hspace{5pt}\phantom{.}$}{v1}
                \fmflabel{$\phantom{.}\hspace{5pt}\frac{\p^6}{\Lambda^5} $}{v2}
\end{fmfgraph*}} +\hspace{10pt}
\parbox{25mm}{\begin{fmfgraph*}(35,50)
	            \fmfleft{i1,i2}
	            \fmfright{o1,o2}
                \fmflabel{$\phi_2$}{i1} 
                \fmflabel{$\phi_1$}{i2} 
                \fmflabel{$\phi_4$}{o1} 
                \fmflabel{$\phi_3$}{o2} 
	            \fmf{dashes,label=$\psi$}{v1,v2}
                \fmf{plain}{i1,v1}
                \fmf{plain}{i2,v2}
                \fmf{plain}{v1,o1}
                \fmf{plain}{v2,o2}
                \fmfdot{v1}
                \fmfdot{v2}
\end{fmfgraph*}}  \hspace{-10pt}+\hspace{10pt}
\parbox{25mm}{\begin{fmfgraph*}(35,50)
	            \fmfleft{i1,i2}
	            \fmfright{o1,o2}
                \fmflabel{$\phi_2$}{i1} 
                \fmflabel{$\phi_1$}{i2} 
                \fmflabel{$\phi_3$}{o1} 
                \fmflabel{$\phi_4$}{o2} 
	            \fmf{dashes,label=$\psi$}{v1,v2}
                \fmf{plain}{i1,v1}
                \fmf{plain}{i2,v2}
                \fmf{plain}{v1,o1}
                \fmf{plain}{v2,o2}
                \fmfdot{v1}
                \fmfdot{v2}
\end{fmfgraph*}} \hspace{-20pt} = \hspace{5pt}0
\end{fmffile}
\end{center}
\caption{Tree-level diagrams contributing to the 4-point function
$\langle \phi^4 \rangle$ at the scale $\Lambda$.}
\end{figure}

The contribution from the first diagram in Fig.1 is given by
\ba
\label{M1}
\mathcal{M}^{(1)}_{\langle \phi^4 \rangle}=-\frac{8 i}{\Lambda^{10}}\,  \((p_1\cdot p_2)^5+2\leftrightarrow 3+ 2\leftrightarrow 4 \)\ \delta^{(4)}(\sum p_j) \,,
\ea
where $p_a$ is the momentum of each external $\phi$ particle, $a=1,\ldots, 4$, and the external legs are computed on-shell such that $p_a^2=0$.
The three last diagrams rely on the exchange of a virtual $\psi$ with $\<\psi^2\>_{p}=-i p^{-2}$, giving the following contribution
\ba
\mathcal{M}^{(n)}_{\langle \phi^4 \rangle}=-\frac{4}{\Lambda^{10}}\,
\, \( (p_1\cdot p_n)^4\<\psi^2\>_{p_1+p_n}(p_1+p_n)^4\)\ \delta^{(4)}(\sum_j p_j) \ ,\  \ \ n=2,3,4\,.
\ea
Computing the external legs on-shell such that $(p_1+p_n)^2=2 p_1 \cdot p_n$, the sum of all diagrams is manifestly zero\footnote{It is easy to check that the same results hold even if the external legs are taken off-shell in which case \eqref{M1} needs to be modified accordingly.},
\ba
\mathcal{M}_{\langle \phi^4 \rangle}=\sum_{n=1}^4 \mathcal{M}^{(n)}_{\langle \phi^4 \rangle}=0\,.
\ea
The same remains true for any $n$-point function $\<\phi^n\>$, (we could also consider here $n$-point functions with external $\psi$ but such diagrams cancel automatically since $\Box \psi$ vanishes on-shell.)
Now, the fact that this theory is well-defined and does not propagate any processes that violate unitarity should not come as a surprise as the cubic term in \eqref{toy} simply indicates that we were not dealing with the properly diagonalized variables, at least around an arbitrary background. Instead, if we were to perform the well-defined, all orders invertible field redefinition,
\ba
\label{fieldredef}
\psi=\bar \psi -\frac{1}{\Lambda^5}(\p_\alpha \p_\beta \bar \phi)^2\,, \hspace{20pt}\phi=\bar \phi \,,
\ea
one would simply uncover a free theory,
\ba
\mathcal{L}=\frac12 \bar \phi \Box \bar \phi+\frac12 \bar \psi \Box \bar \psi\,,
\ea
which is manifestly healthy. One might worry that field redefinitions as \eqref{fieldredef} are prohibited as they change the order of the field. Such an argument would of course contradict the fact that the S-matrix is independent of the choice of variable.  Fortunately,
the result with or without field redefinition is of course  just the same, as we have explicitly shown here, one simply needs to work harder to derive the correct physical outcomes. Furthermore as we shall show below perturbation theory remains under control even at scales above $\Lambda$ and even when working with the original variables $\psi,\phi$. To complete the analogy with massive gravity, the $\phi,\psi$ variables represent the helicity-0 and helicity-2 modes defined in the helicity decomposition, whereas  $\bar \phi,\bar \psi$ represent their cousins in the \stu decomposition.
\\

To make the unitarity of the theory manifest in this example, we have not redefined the field $\phi$ itself, but have instead diagonalized the field $\psi$ by a linear shift which depends on $\phi$. This distinction is crucial. If we were dealing with a single field theory, such a situation would not have been possible and any interactions at the scale $\Lambda$ would always have signaled the breakdown of perturbation theory regardless of the ability to redefine them away. \\

To give an example of {\bf what we are not doing} and where concerns about field redefinitions containing derivatives are well founded, consider a single field model which takes the form
\ba
\mathcal{L} = \frac 12 \chi \Box \chi + \frac{1}{\Lambda^3} \chi \left( \Box \chi\right)^2 + \frac{1}{2\Lambda^6} \chi \Box \chi \Box \( \chi \Box \chi \)
\ea
This model certainly contains more solutions than allowed for a single field theory with a well-defined Cauchy problem. To see this, expand the theory around a background solution $\chi=\mu$, where $\mu \ll \Lambda$ is a constant. To quadratic order the action is
\ba
{\mathcal L}^{(2)} = \frac{1}{2} \delta \chi \Box \delta \chi + \frac{1}{\Lambda^3} \mu \delta \chi  \Box^2 \delta \chi + \frac{1}{2\Lambda^6} \mu^2 \delta \chi \Box^3 \delta \chi \,,
\ea
hence the equation of motion for the perturbations is
\ba
\Box \(1 + \frac{ \mu}{\Lambda^3} \Box \)^2 \delta \chi=0\,,
\ea
which exhibits a healthy massless mode and a ghostly mode with mass $m^2 = \Lambda^3/\mu$. This is just a special case of the general result made apparent through a standard Ostrogradsky analysis.

However, we may also choose to perform the field redefinition $\bar{\chi} = \chi + \chi( \Box \chi)/\Lambda^3$ and see that the theory appears to become free
\ba
\mathcal {L} = \frac{1}{2} \bar \chi \Box \bar \chi\,.
\ea
However this is clearly fake. The original theory has more solutions than the redefined one. The original theory has a ghostly pole around the background solution $\chi = \bar \chi = \mu$. In terms of $\bar \chi$ the ghost seems to have disappeared, but in reality the field redefinition has simply disguised it. The field redefinition becomes ill-defined at the scale $\Lambda$, \ie when $\chi \sim \Lambda$ and $\partial \sim \Lambda$ since at that point perturbation theory breaks down since $\chi( \Box \chi)/\Lambda^3 \sim \chi$. This means that it is not possible express $\chi$ in terms of $\bar{\chi}$ via a convergent perturbation expansion. This problem becomes manifest if the coupling to an external source is through $\chi$. Thus whilst below the scale $\Lambda$ we may work with the variable $\bar \chi$, the ghost reappears at the scale $\Lambda$.

For massive gravity the situation is completely different. The number of degrees of freedom even in the presence of matter remains 5 in both variables. Thus the field redefinition between the helicity and \stu variables could never have disguised any degrees of freedom. This is backed up by the fact that the field redefinition remains under perturbative control below the scale $\Lambda_3$. We shall demonstrate this in the next section, but doing so let us first end this section by commenting on order by order field redefinitions.  \\

Let us consider a slightly different example. Suppose a theory in which the scalar sector to third order in an expansion in $\chi$ takes the form
\ba
\mathcal{L} = \frac12 \chi \Box \chi + \frac{1}{\Lambda^3} \chi \( \Box \chi\)^2 + {\cal O}(\chi^4)\, .
\ea
Can we say that there exists a ghost at the scale $\Lambda$? Here the answer is clearly no, precisely because to the same order the cubic term is removable by the same field redefinition $\bar{\chi} = \chi + \chi( \Box \chi)/\Lambda^3$. This means that to order $(E/\Lambda)^3$ there are no interactions. Whilst this term does give unitarity violation at higher order $(E/\Lambda)^6$ we cannot say that there is truly unitarity violation at this scale until we have computed the Lagrangian to sufficiently high order in the field expansion so that all terms that can contribute at the same order are accounted form. In other words, by neglecting interactions that can contribute at the same order in an expansion in $E/\Lambda$ we might wrongly conclude the existence of a ghost or a breakdown in perturbation theory where there is not one. \\

\subsection{How to have higher derivatives and evade ghosts}

Ostrogradsky's famous argument, adopted to the case considered  here,  is that higher derivative interactions inevitably lead to ghost is predicated on the observation that theories with higher derivatives require additional initial data to specify the Cauchy problem. This extra initial data needed is always found to be precisely the initial data defining the ghost degree of freedom. It is easy to see that the model \eqref{toy} has a well-defined Cauchy problem {\it without} the need to specify additional initial data. This is why there is no ghost in this model despite the appearance of higher derivatives in the Lagrangian. \\

Let us derive the equations of motion from \eqref{toy}. There are two equations:
\ba
\label{eoms1}
&&\mathcal{E}_{\phi}=\Box \phi+\frac{2}{\Lambda^5}\p_\mu\p_\nu\left[\p^{\mu}\p^{\nu} \phi \left(\Box \psi+ \frac{1}{\Lambda^5} \Box ((\p_\alpha \p_\beta \phi)^2)\right) \right] =0 \\ \nonumber
&& \mathcal{E}_{\psi}=\Box \psi+\frac{1}{\Lambda^5} \Box ((\p_\alpha \p_\beta \phi)^2)=0.
\ea
These equations appear  to have  higher derivatives,  suggesting that they contain ghosts.
But it is easy to see that this
is not the case.  The first equation in (\ref {eoms1}), with the account of the second one,
reduces to a trivial  second order equation. More formally speaking, taking a simple combination
these two equations give
\ba
\mathcal{E}_{\phi}-\frac{2}{\Lambda^5}\p_\mu\p_\nu\left[\p^\mu\p^\nu \phi \, \mathcal{E}_{\psi} \right] =\Box \phi= 0\,.
\ea
The key point is that this equation is a standard second order differential equation for $\phi$, and so requires just the usual two pieces of initial data. The remaining equation to solve, $\mathcal{E}_\psi=0$
does contain higher derivatives in $\phi$, but these are already fixed by the solution of the first equation\footnote{Naturally, we are considering a situation with no discontinuity, so the equations of motion are satisfied at least in an infinitesimal region prior where the Cauchy surface is taken.}.
Thus there are only two remaining pieces of initial data needed, giving a total of four: The same as in a theory of two free scalar fields. So even without the use of any field redefinition, and even at the nonlinear level, since the theory requires the same amount of initial data as usual, there can be no ghosts.

The above observation also explains why there is no difficulty in performing the field redefinition \eqref{fieldredef},  even though this redefinition requires double derivatives of $\phi$. This is because all double derivatives are completely fixed by the usual Cauchy data along with the well-defined equation of motion for $\phi$.

\subsection{Coupling to sources}

\label{couplingtosources}

The only subtlety in the previous arguments has to do with the coupling to sources, be they external or additional fields. In the case of massive gravity to be treated in what follows this is concerned with the issue of how to couple with matter\footnote{We point out however that the apparent presence of a ghost in \cite{slavapaper1} and \cite{Folkerts:2011ev} is not attributed to the presence of matter.}. The essential point here is that {\it a generic coupling to sources could reintroduce the ghosts/perturbation theory problems at the scale $\Lambda$.} For instance this can be seen easily by naively coupling to an external source by the addition of a term in the Lagrangian of the form
\ba
\mathcal{L}_{\rm source} = J_{\phi} \phi + J_{\psi} \psi\,.
\ea
This coupling introduces problems at the scale $\Lambda$ as can be seen by recomputing the equation of motion for $\phi$ in the presence of the source
\ba
\mathcal{E}_{\phi}-\frac{2}{\Lambda^5}\p_\mu \p_\nu\left[\p^\mu \p^\nu \phi \, \mathcal{E}_{\psi} \right] =\Box \phi+J_{\phi}-\frac{2}{\Lambda^5}\p_\mu \p_\nu\left[\p^\mu \p^\nu\phi \, J_{\psi} \right]  = 0\,.
\ea
This equation now explicitly includes higher derivatives at the scale $\Lambda$ and the previously vanquished Ostrogradsky ghost reappears. \\

The resolution of this problem is to be more careful in the choice of couplings to sources. At the classical level, we can choose to couple how we see fit, at the quantum level we must ensure that the choice made classically is preserved under quantum corrections. It is easy to see that this can be achieved in our toy scalar example. For instance for an external source, instead of the previous coupling we choose to couple as
\ba
\mathcal{L}_{\rm source} = \bar{J}_{\phi} \phi + \bar J_{\psi} \left( \psi +\frac{1}{\Lambda^5} (\p_\alpha \p_\beta \phi)^2 \right)\,,
\ea
we easily see that the equation of motion for $\phi$ now becomes simply
\ba
\mathcal{E}_{\phi}-\frac{2}{\Lambda^5}\p_\mu \p_\nu \left[\p^\mu \p^\nu  \phi \, \mathcal{E}_{\psi} \right] =\Box \phi+\bar{J}_{\phi}  = 0\,,
\ea
and we recover the result that the Ostrogradsky ghost is vanquished. All of this is made manifest by the field redefinition, which shows that the theory is equivalent to a free theory with external sources
\ba
\mathcal{L}=\frac12 \bar \phi \Box \bar \phi+\frac12 \bar \psi \Box \bar \psi +  \bar{J}_{\phi} \bar \phi + \bar J_{\psi} \bar \psi .
\ea
But the point is, since this is equivalent to a free theory, this form of coupling to $\phi$ and $\psi$, even though apparently fine tuned in the original variables, is nevertheless preserved to all orders under quantum corrections! For instance, even if the external sources $\bar{J}_{\phi}$ and $\bar J_{\psi} $ are made explicit functions of matter fields, there are no interactions at the scale $\Lambda$ to reintroduce the pathological couplings at any order in a loop expansion. Although this toy model is special, it is clear that it is always possible to:
\begin{enumerate}
\item Choose couplings to sources which preserves the consistency of perturbation theory and absence of ghosts at the scale $\Lambda$;
\item This choice can be made so that it is {preserved under quantum corrections.}
\end{enumerate}

In the case of massive gravity the implications of this are that it is possible to
{\bf choose consistent couplings to matter which preserve the absence of ghosts/strong coupling up to  the scale $\Lambda_3$}. This choice of couplings can be made so that it is {\bf preserved under quantum corrections}. We should hardly be surprised by this result. It is well understood that in GR the couplings to matter must respect diffeomorphism invariance otherwise ghosts will appear at a scale well below the Planck scale. Thus it is no surprise that we have to impose similar restrictions in massive gravity even though diffeomorphism invariance is broken. Of course in GR this situation is safer since in the absence of an anomaly, diffeomorphism invariance guarantees that the matter couples in the correct way to all orders in quantum corrections. In massive gravity, at present we can guarantee this at least within the regime of validity of the effective field theory. It is clear that all the current formalisms to describe massive gravity do not yet present all its special properties in a manifest way. It is likely that there exists another formulation of the allowed ghost-free massive gravity models that will make the consistency of coupling to matter explicit.

\subsection{Perturbation theory is well-defined}

Finally we conclude this scalar field toy-model by showing why perturbation theory is well-defined here even when the field redefinition is large.
The redefinition \eqref{fieldredef} may still give a lingering doubt that since the second term becomes as large as the first at energy scales $\partial \sim \Lambda$ and $\phi \sim \psi \sim \Lambda$, then perturbation theory is breaking down. This was the argument suggested in \cite{Folkerts:2011ev} that the perturbation theory for the spin-2 field was breaking down, applied in the present toy-model. However, this is not the condition for the validity of perturbation theory. Rather the condition is that the perturbative expansion
converges (at least is an asymptotic series, although in the present case this is not a concern). The second term may be as large or larger than the first as long as the higher order terms become sufficiently small that the whole perturbative expansion converges. Fortunately this is trivial to see here, not only does the expansion converge, it terminates. Recognizing that the expansion parameter is $\Lambda^{-5}$, \ie $(E/\Lambda)^5$ we may express the general solutions of the equations of motion \eqref{eoms1} in the form
\ba
&&\phi = \sum_{n=0}^{\infty} \frac{1}{\Lambda^{5n}}\phi^{(n)} \\
&&\psi =  \sum_{n=0}^{\infty} \frac{1}{\Lambda^{5n}}\psi^{(n)}\,.
\ea
It is easy to see that $\phi^{(n)}=0$ for $n \ge 1$ and $\psi^{(n)}=0$ for $n \ge 2$ and that the only nonzero terms are
\ba
\phi^{(0)} &=& \phi_{h}-\Box^{-1} \bar J_{\phi} \\
\psi^{(0)} &=& \psi_{h}-\Box^{-1} \bar J_{\psi}  \\
\psi^{(1)} &=&  -(\p_\alpha \p_\beta \phi^{(0)})^2\,,
\ea
where $\phi_h$ and $\psi_h$ are homogeneous solutions $\Box \phi_h=\Box \psi_h=0$.

Thus whilst it may be true that above the scale $\Lambda$ the second order perturbation may be larger than the first, \ie $\psi^{(1)} \gg \psi^{(0)}$, perturbation theory is still well-defined because the third and all higher perturbations are zero. Thus the perturbative expansion always converges and is moreover exact at first order. In the full massive gravity theory, this argument can be used to see that perturbation theory only breaks down, \ie that the perturbative expansion fails to converge, only once we hit the scale $\Lambda_3$ where the helicity-0 mode becomes strongly coupled. In fact in massive gravity, the situation is even clearer. At the same scales at which the second order perturbation becomes larger than the first it is still true that $h_{\mu\nu}/\mpl  \ll 1$ and this explains why the perturbative series is still well under control.

\vspace{20pt}

\section{Ghost-free Massive Gravity}
\subsection{The model}

In this paper we will consider only the two parameter family of the ghost-free models proposed in \cite{deRham:2010kj} and proven ghost free in \cite{Hassan:2011hr}. This is what we mean by `Ghost-free Massive Gravity'. We consider the metric $g\mn$,  and in what follows, we use the notation where square brackets $[\ldots]$ represent the trace of a tensor contracted using the Minkowski
metric, {\it e.g.} $[\mathcal{K}]=\eta^{\mu\nu}\mathcal{K}_{\mu\nu}$ and $[\mathcal{K}^2]=\eta^{\alpha \beta}\eta^{\mu\nu}\mathcal{K}_{\alpha\mu}\mathcal{K}_{\beta\nu}$, while angle brackets $\langle \ldots \rangle$ represent the trace with respect to the physical metric $g\mn$, so that $\langle H \rangle = g^{\mu\nu}H_{\mu\nu}$ and $\langle H^2 \rangle = g^{\alpha \beta}g^{\mu\nu}H_{\alpha\mu}H_{\beta\nu}$. \\

The action for the allowed interacting theories of massive spin-2 fields written in a diffeomorphism invariant way are then of the form \cite{deRham:2010kj},
\label{sec:model}
\ba
\label{L2}
\mathcal{L}=2 \mpl^2\sqrt{-g} R+\mathcal{L}_m\,,
\ea
where $R$ is the scalar curvature and the mass term is given by
\ba
\label{Lm}
\mathcal{L}_m=2 m^2\mpl^2 \sqrt{-g}\(\mathcal{L}^{(2)}_{\rm der}(\K) + \alpha_{3} \mathcal{L}^{(3)}_{\rm der}(\K) +
\alpha_{4}  \mathcal{L}^{(4)}_{\rm der}(\K)\)
\ea
with the interaction terms are given by
\ba
\label{L2der0}
\mathcal{L}^{(2)}_{\rm der} &=&[ \K ]^2- [\K^2] \,,\nn\\
\label{L3der}
\mathcal{L}^{(3)}_{\rm der}&=&[ \K ]^3-3 [ \K ] [\K^2] +2[\K^3]\,,\\
\label{L4der}
\mathcal{L}^{(4)}_{\rm der}&=&[ \K ]^4-6 [ \K^2] [ \K ]^2+8[ \K^3]
[\K]+3[ \K^2 ]^2-6[\K^4]\,,\nn
\ea
and $\mathcal{K}$ is defined via $2\mathcal{K}^{\mu}_{\nu}- \mathcal{K}^{\mu}_{\alpha}\mathcal{K}^{\alpha}_{\nu} = H^{\mu}_{\nu}$ or equivalently $\K^\mu_\nu = \delta^\mu_\nu -
\sqrt{\partial^\mu \phi^a \partial_\nu\phi^b\eta_{ab}}$,
with
\ba
H_{\mu\nu} = g_{\mu\nu} - \eta_{ab} \partial_{\mu} \phi^a  \partial_{\nu} \phi^b\,.
\ea
The $\phi^a$'s are four \stu fields introduced to provide a diffeomorphism invariant formalism.
Throughout the rest of the paper we shall work in unitary gauge for which $\phi^a=x^a$ and
so $H_{\mu\nu}=g_{\mu\nu}-\eta_{\mu\nu}=h_{\mu\nu}/\mpl $. \\

The allowed mass terms derived in \cite{deRham:2010kj}  can be shown to arise as the expansion of the determinant $\sqrt{-g}\det[g\mupn+\lambda \K\mupn]$ to fourth order in
$\lambda$ \cite{Koyama,Nieuwenhuizen,Hassan:2011vm}. In particular, it is easy to show that up to a total derivative, a linear combination of the above three interaction terms in \eqref{L2der0}, is equivalent to a single term $[\K]$ in combination with a cosmological constant. This is referred to as the minimal model in \cite{Hassan:2011vm}.

With these terms it was shown that no ghosts are present in the models, neither in the decoupling limit \cite{deRham:2010kj,deRham:2010ik}, nor in the full theory by performing a Hamiltonian analysis via the ADM formulation \cite{deRham:2010kj,Hassan:2011hr}, nor beyond the decoupling limit in the \stu formulation \cite{deRham:2011rn}.

\subsection{Helicity decomposition}

The essence of the helicity decomposition is the observation that at physical momenta $k \gg m$ the spin-2 representation may naturally be decomposed into two helicity-2 modes, two helicity-1 modes, and a single helicity-0 mode. The decomposition that serves to diagonalize the quadratic order kinetic term is, written in terms of the metric $g_{\mu\nu}=\eta_{\mu\nu}+h_{\mu\nu}/\mpl$ where
\ba
\label{helicity_dec}
h_{\mu\nu}=\tilde{h}_{\mu\nu} + \frac 1{2m} \partial_{(\mu} A_{\nu)} +\frac{1}{3}\( \frac{1}{m^2}\Pi\mn +\frac 12  \pi \, \eta_{\mu\nu}\)\,,
\ea
with $\Pi\mn=\p_\mu \p_\nu \pi$. (Here $\tilde{h}_{\mu\nu}$ is what was earlier referred to as $h_{\mu\nu}^\hel $).
An important point to note, is that whilst the last term is subdominant to the one that precedes it, it is crucial to include it in order to diagonalize the kinetic term. Without it there will be a kinetic mixing between $\pi$ and $\tilde{h}_{\mu\nu}$. The fact that subdominant terms must be included to remove the kinetic mixing will play an important role in the following discussion. \\

This decomposition can for example be useful in considering scattering amplitudes, since it correctly identifies the asymptotic  helicity-2 mode $\tilde{h}_{\mu\nu}$, two helicity-1 modes $A_{\mu}$, helicity-0 mode $\pi$. The decomposition also respects linearized diffeomorphisms and a U(1) symmetry which guarantees the correct counting of degrees of freedom of the free theory. \\

As in the \stu decomposition, the $1/m^2$ in the normalization of the helicity-0 mode implies its interactions come in at a different scale than those of the helicity-2. As such we can take a decoupling limit in which $\mpl  \rightarrow \infty $ and $m \rightarrow 0$ keeping combinations of the form $\Lambda_n = (\mpl m^{n-1})^{1/n}$ fixed to isolate the dominant interactions.

To start with, it will be useful to perform an expansion in $h\mn$, such that we have,
\ba
\mathcal{L}_m=-\frac{m^2}{2}(h\mn^2-h^2)
+\frac{m^2}{\mpl} \(k_1h\mn^3+k_2h h\mn^2+k_3h^3\)
+\mathcal{O}(h^4)\,,
\ea
with
\ba
\label{k's}
k_1=-(k_2+k_3)\,, \hspace{15pt}
k_2=-(1+3\alpha_3/4)\hspace{10pt}{\rm and}\hspace{10pt}
k_3=(1+\alpha_3)/4 \,,
\ea
and all indices are raised with $\eta\mn$. Similarly, the Einstein-Hilbert (EH) term, is of the form
\ba
2\mpl^2\sqrt{-g} R&=&\frac 1{2} h^{\mu\nu} \Ein^{\alpha\beta}\mn h_{\alpha\beta}
+\frac{1}{2\mpl}h^{\mu\nu} \Big[\p_\mu h^\alpha_{[\beta}\p_\nu h^\beta_{\alpha]}
-2 \p_\alpha h^\beta_{[\beta}\p^\alpha h_{\mu]\nu}
+2 \p_\alpha h_{\mu(\nu}\p^\beta h_{\beta)}^\alpha\\
&-&4\p_{(\alpha}h^\alpha_\mu\p_\beta h^\beta_{\nu)}
-\frac 12 \eta\mn \(\p_\alpha h^\gamma_{[\beta}\p^\alpha h^\beta_{\gamma]}+2 \p_\alpha h^\gamma_{(\gamma}\p_{\beta)} h^{\alpha\beta}
+4 (\p_\alpha h^\alpha_\beta)^2\)\Big]
+\mathcal{O}(h^4)\nn\,,
\ea
where $\hat{\mathcal{E}}$ is the linearized Einstein tensor,
\ba
\Ein^{\alpha\beta}\mn h_{\alpha\beta}=\Box h\mn-\p_\alpha \p_{(\mu}h^\alpha_{\nu)}+\p_\mu\p_\nu h-\eta\mn (\Box h-\p_\alpha\p_\beta h^{\alpha\beta})\,,
\ea and we define $(a,b)=ab+ba$ while $[a,b]=ab-ba$. In terms of the helicity decomposition \eqref{helicity_dec}, the full quadratic action is then
\ba
\label{L2tot}
\mathcal{L}^{(2)}&=&\frac 12 \tilde h^{\mu\nu}\Ein^{\alpha\beta}\mn \tilde h_{\alpha\beta}-\frac 12 F\mn^2+\frac 1{12}\pi \Box \pi\\
&-&m A_\mu \(\p_\nu \tilde h^{\mu\nu}-\p_\mu \tilde h\)+\frac{m}{2}\pi \p_\mu A^\mu\nn\\
&-&\frac 12 m^2\(\tilde h\mn^2-\tilde h^2\)+\frac12 m^2 \pi \tilde h+\frac 16 m^2 \pi^2\nn\,.
\ea

\section{Relevant interactions at the scale $\Lambda_5$}
\label{sec:L5}

In this section we shall derive the decoupling limit obtained keeping the scale $\Lambda_5$ fixed and show that the resulting theory is a ghost-free free theory.
The most dangerous interactions that could arise in a massive theory of gravity occur at the scale $\Lambda_5$. At cubic order, the interactions that arise at the scale $\Lambda_5$ are that of the form $(\p^2 \pi)^3$ arising both from the EH term and the mass term, as well as $\tilde h \p^2 (\p^2\pi) (\p^2 \pi)$ arising solely from the EH term.
This second kind of interactions, play an essential role in the consistency of the theory. The most relevant interactions at cubic order are,
\ba
\label{L3_5}
\mathcal{L}^{(3)}_{\Lambda_5}&=&-\frac{1}{18\Lambda_5^5}\tilde h^{\mu\nu}\(V\mn-\frac 12 V \eta\mn\) \\
&+&\frac{1}{27 \Lambda_5^5}\(\frac{1}{4}\pi V+k_1[\Pi^3]+k_2[\Pi][\Pi^2]+k_3[\Pi]^3\) \,,\nn
\ea
with
\ba
V\mn=\p_\mu \p_\alpha \p_\beta \pi \p_\nu \p^\alpha \p^\beta \pi-\p_\alpha \p_\mu\p_\nu \pi \p^\alpha \Box \pi\,.
\ea
and $V=\eta^{\mu\nu}V_{\mu\nu}$.
As a consistency check, one can see that $V\mn$ satisfies the transverse relation $\p_\mu V\mupn=\frac 12 \p_\nu V$, which is a simple consequence of linearized diffeomorphism invariance. (In the decoupling limit, full diffeomorphism invariance always reduces to the linearized diffeomorphism invariance, but this is true only in the decoupling limit \cite{deRham:2011rn,deRham:2010tw}). One can also check that $V\mn$ is nothing but the Lichnerowicz operator applied on $\Pi^2\mn=\p_\mu \p_\alpha \pi \p_\nu \p^\alpha \pi$,
\ba
V\mn-\frac 12 V\eta\mn=\frac 12\,  \Ein^{\alpha\beta}\mn \,  \Pi^2 _{\alpha\beta}\,,
\ea
which will have important consequences for the consistency of the theory, as we will see below. After integration by parts, we simply have the relation
\ba
\pi V=\frac 12 ([\Pi]^3-[\Pi][\Pi^2])\,.
\ea
Unsurprisingly, the mass term in \eqref{Lm} is precisely such that the last line \eqref{L3_5} is a total derivative,
\ba
\frac{1}{4}\pi V+k_1[\Pi^3]+k_2[\Pi][\Pi^2]+k_3[\Pi]^3
=\frac{3+2\alpha_3}{8}\([\Pi]^3-3 [\Pi][\Pi^2]+2 [\Pi^3]\)\,.
\ea
However, it is not possible, nor is it desirable to cancel the $\tilde h V $ interactions, because they play a crucial role in ensuring the consistency of the theory.
This leads to the cubic vertex at the scale $\Lambda_5$
\ba
\label{L3_5_sim}
\mathcal{L}^{(3)}_{\Lambda_5}=-\frac{1}{36 \Lambda_5^5}\left[\Pi^2 \right]^{\mu\nu}\Ein ^{\alpha\beta}\mn\tilde h_{\alpha\beta}\,.
\ea
The full decoupling theory at the scale $\Lambda_5$ also includes quartic interactions of the form
\ba
\mathcal{L}^{(4)}_{\Lambda_5}&=&\frac{1}{3^4 2^4\Lambda_5^{10}}\, \Pi^2\mn (V^{\mu\nu}-\frac 12 V\eta^{\mu\nu})
=\frac{1}{3^4 2^5\Lambda_5^{10}}\, \Pi^2\mn \, \Ein_{\alpha\beta}^{\mu\nu} \,  \Pi^2{} ^{\alpha\beta}  \,.
\ea
Putting this together, the exact decoupling theory obtained in the limit $\mpl  \rightarrow \infty$ keeping $\Lambda_5$ fixed is
\ba
\label{dec5}
\mathcal{L}^{\rm dec}_{\Lambda_5} =
\frac12 \(\tilde h^{\mu\nu}-\frac{\Pi^2{}^{\mu\nu}}{36 \Lambda_5^5}\) \Ein^{\alpha \beta}\mn
 \(\tilde h_{\alpha\beta}-\frac{\Pi^2_{\alpha \beta}}{36 \Lambda_5^5}\)
-\frac 18 F\mn^2 +\frac 1{12} \pi \Box \pi\,.
\ea
We shall analyze the physics of this action below.

\subsection{Triviality of the S-matrix in the $\Lambda_5$ theory}
A cursory glance at this action \eqref{dec5} would seem to imply the presence of strong coupling, or unitarity violation at the scale $\Lambda_5$. However it is easy to see that this is not the case, since there is something remarkably special about this theory that is not apparent at first. To see what this is, let us calculate scattering amplitudes. The cubic vertex $\mathcal{L}^{(3)}_{\Lambda_5}$ in \eqref{L3_5_sim} can be used in diagrams but only if $\tilde h$ is computed off-shell (no external leg in $\tilde h$). There is therefore no non-trivial 3-point function at the scale $\Lambda_5$\footnote{If we were to compute the 3-point scattering amplitude on top of a background, for which we could take $\tilde h$ off-shell, it is true that this vertex could give rise to a non-trivial contribution to the 3-point function, but it would be canceled by the contribution from the matter.}.
For the 4-point function, on the other hand, the situation is different and let us calculate the $2 \rightarrow 2$ scattering amplitude between four helicity-0 modes at tree level. This amplitude receives a contribution from 4-point interaction encoded in $\mathcal{L}^{(4)}_{\Lambda_5}$. Taking this term alone one would find that the scattering amplitude violates perturbative unitarity at the scale $\Lambda_5$. Thus, if the decoupling theory only included the term $\mathcal{L}^{(4)}_{\Lambda_5}$, it would certainly be true that the theory would violate tree level unitarity at the scale $\Lambda_5$. However, there are 3 other Feynman diagrams that contribute at tree level at the same order corresponding to s-channel, t-channel and u-channel exchange of a virtual helicity-2 mode. Individually each of these diagrams also violates perturbative unitarity at the scale $\Lambda_5$. However the sum of all the diagrams which arise at the same order $1/\Lambda_5^{10}$ is easily shown to be zero, in just the same way as was performed in the preamble scalar field toy-model (see Fig.1). \\

Since only the total amplitude can be measured, and the total scattering amplitude is zero, we find that there is no unitarity violation at the scale $\Lambda_5$. This in turn implies there are no new degrees of freedom, or ghosts at this scale. This argument can easily be extended to all $n$-point functions and to all loops,
and one may confirm by direct although laborious calculation that the S-matrix is trivial
\ba
\langle f |\, \hat{S}-\hat{\mathbb{I}}\,  | i \rangle = 0\,.
\ea
There is fortunately a much simpler way to see the triviality of the S-matrix. By performing the well-defined, invertible field redefinition
\ba
\label{redef1}
\tilde{h}_{\mu\nu} = \bar{h}_{\mu\nu} + \frac{1}{36\Lambda_5^{5}}\Pi\mn^2\,,
\ea
we can easily see that the decoupling theory reduces to
\ba
\mathcal{L}^{\rm dec}_{\Lambda_5} =
\frac12 \bar h^{\mu\nu} \Ein^{\alpha \beta}\mn
 \bar h_{\alpha\beta}-\frac 18 F\mn^2 +\frac 1{12} \pi \Box \pi\,,
\ea
where the free kinetic term is now written in terms of $\bar{h}_{\mu\nu}$. Thus since the theory is free in this decoupling limit we can see easily why the S-matrix turned out to be trivial to all orders. \\

One might worry that such redefinitions are not appropriate since they mix orders of the spin-2 field and hence imply that at least one of the degrees of freedom is strongly coupled. However, this is clearly incorrect, since we did not need to perform this field redefinition to see that the S-matrix was trivial. It was possible to see it in the original variables. Since the S-matrix is trivial, all the degrees of freedom are not only weakly coupled at these scales, they are precisely free in the limit. This fact is true regardless of the variables used. The S-matrix is of course only an on-shell quantity, however we shall see below that perturbation theory is under control even off-shell.\\

Although not necessary, the field redefinition \eqref{redef1} is desirable to perform since it diagonalizes the kinetic mixing of the helicity-2 and -0 modes at cubic order. Remarkably it is precisely the same contribution we derive using the \stu decomposition. However, we did not have to invoke the \stu decomposition nor diffeomorphism invariance to see this. It follows simply from diagonalization of the kinetic term. Furthermore since the redefinition is quadratic in the fields, it does not change the asymptotic states, and so both $\tilde{h}_{\mu\nu}$ and $\bar{h}_{\mu\nu}$ deserve to be called the helicity-2 mode. Note also that the additional piece seen in  \eqref{redef1} is a modification of the subleading part of the decomposition since it is down by $m^2/k^2$ relative to the leading order helicity-0 contribution $\p^2 \pi/m^2$ when $\pi \sim \p \sim \Lambda_5$.

\subsection{Absence of ghosts nonlinearly or around arbitrary backgrounds}

The previous arguments guarantee the consistency of the theory defined perturbatively around $h_{\mu\nu}=0$. However, sometimes it is possible for ghosts to arise when expanded around an arbitrary background. Again, if we neglect the $\tilde{h} V$ interaction this is true of the theory \eqref{dec5}. However, when all interactions are included it is easy to show that this is not the case. To see this, let us derive the full equations of motion, in the original helicity variables. In an arbitrary gauge the equations of motion that follow from \eqref{dec5} are
\ba
&&\mathcal{E}_{\pi}= \Box \pi - \frac{1}{3 \Lambda_5^5}\partial^{\omega} \partial^{\nu} \left[ \partial^{\mu} \partial_{\omega} \pi
{\Ein}^{\alpha \beta}_{\mu\nu} \(\tilde{h}_{\alpha \beta} - \frac{1}{36\Lambda_5^{5} }\Pi^2_{\alpha\beta}\) \right] =0 \, , \\
&& \mathcal{E}_{A_{\mu}}=\partial^{\mu} F_{\mu\nu} =0 \, , \\
\label{tildeh}
&&\mathcal{E}_{h_{\mu \nu}}={\Ein}^{\alpha \beta}_{\mu\nu} \(\tilde{h}_{\alpha \beta} - \frac{1}{36\Lambda_5^{5} }\Pi^2_{\alpha\beta}\)=0\,.
\ea
As usual these equations are seemingly higher derivative, and so a cursory analysis would seem to suggest the presence of a ghost. However, again this can be shown not to be the case. As before taking the combination
\ba
\mathcal{E}_{\pi}+ \frac{1}{3 \Lambda_5^5}\partial^{\omega} \partial^{\nu} \left[ \partial^{\mu} \partial_{\omega} \pi \,  \mathcal{E}_{h_{\mu \nu}} \right] = \Box \pi =0.
\ea
This gives a well-defined equation of motion for $\pi$ whose solution, once known, can be substituted back into the equation \eqref{tildeh} for $\tilde h\mn$. This analysis is completely analogous to what is performed in the \stu language beyond the decoupling limit, see Ref.~\cite{deRham:2011rn}.

The number of degrees of freedom is one half the total amount of initial data needed to solve these equations. As usual the vector part gives two helicity-1 modes. Crucially the equation for the helicity-0 is second order and so gives rise to only one independent degree of freedom (two pieces of initial data). Finally, although the right hand side of the equation for the helicity-2 mode \eqref{tildeh} contains higher derivatives, the helicity-0 mode is already determined by its own well-defined equation. Thus the only remaining initial data to specify is that for $\tilde{h}_{\mu\nu}$. As usual, the gauge condition removes 4 degrees of freedom, but if say we choose de Donder gauge, since it does not fix all the freedom, we can use the remaining
non-uniqueness to reduce down to two independent polarizations which describe the real
helicity-2 component. \\

All of the above is trivial once we work with the field redefined coordinates for which the equations of motion take the form
\ba
&&\Box \pi =0 \, ,\nn  \\
&& \partial^{\mu} F_{\mu\nu} =0\,,  \\ \nn
&&{\Ein}^{\alpha \beta}_{\mu\nu} \bar{h}_{\alpha \beta} = 0\, .
\ea
It is however important to make clear that the field redefinition does not disguise any degrees of freedom, which is why we have labored the argument. The counting can be performed correctly without performing any field redefinitions. \\

Note that again had we discarded the $\tilde{h}\p^2 (\p^2 \pi)^2$ interaction, as was done in  \cite{Folkerts:2011ev}, the equation of motion for the helicity-0 mode would have been higher order, and it would have been necessary to supply more initial data to solve it. This would indeed indicate the presence of a ghost at the scale $\Lambda_5$. Only once all the interaction terms are taken into account can we correctly diagnose the absence of a ghost.

\subsection{Coupling to matter}
\label{sec:Matter}

\subsubsection{A first look at external sources}

Let us now come to the coupling to matter. We already know from our simple two scalar toy model that this deserves some care. For instance, let us suppose as a first attempt that we just couple to an external source $S^{\mu\nu}$ which we shall not assume to be conserved. This is important since diffeomorphism invariance is broken by the mass term, there is no reason to assume that the external source for a massive spin-2 field is conserved. Adding the term
\ba
\Delta \mathcal{L}_{\rm matter} = \frac{1}{2} h_{\mu \nu} S^{\mu\nu}\,,
\ea
and taking the $\Lambda_5$ decoupling limit modifies the equations of motions to
\ba
&&\hspace{-10pt}\mathcal{E}_{\pi}= \Box \pi - \frac{1}{3 \Lambda_5^5}\p^{\omega} \p^{\nu} \left[ \p^{\mu} \p_{\omega} \pi
{\Ein}^{\alpha \beta}_{\mu\nu} \(\tilde{h}_{\alpha \beta} - \frac{1}{36\Lambda_5^{5} }\Pi^2_{\alpha\beta}\)
\right]+\frac{1}{m^2} \partial^{\mu} \partial^{\nu}S_{\mu\nu} + \frac{1}{2} S=0  \\
\label{vector_mat}
&& \hspace{-10pt} \mathcal{E}_{A_{\nu}}=\partial^{\mu} F_{\mu\nu} +\frac{1}{2 m} \partial^{\mu}S_{\mu\nu}=\partial^{\mu} F_{\mu\nu} =0 \, , \\
\label{tildeh_mat}
&& \hspace{-10pt} \mathcal{E}_{h_{\mu \nu}}={\Ein}^{\alpha \beta}_{\mu\nu} \(\tilde{h}_{\alpha \beta} - \frac{1}{36\Lambda_5^{5} }\Pi^2_{\alpha\beta}\)+\frac{1}{2 }S_{\mu\nu}=0\,.
\ea
The absence of a source in the helicity-1 equation need a qualification:
In the decoupling limit $m \rightarrow 0$ the source takes the form $S_{\mu\nu} = S^{(0)}_{\mu\nu} +m S^{(1)}_{\mu\nu} +m^2 S^{(2)}_{\mu\nu}+... $,
where $S^{(0)}_{\mu\nu}$ is conserved, $\partial^\mu S^{(0)}_{\mu\nu} =0$,
while in general $S^{(1,2)}_{\mu\nu}$ is not. However, we will show below that $ \partial^\nu S^{(1)}_{\mu\nu} \sim m$, which is a sufficient condition to guarantee that the source in
the helicity-1 equation $\mathcal{E}_{A_{\nu}}$ vanishes in the decoupling limit,
and hence trivially satisfies the Bianchi identity to that order.
In what follows, for all nonconserved
currents on the r.h.s. of the helicity-1 equation this decomposition will be implied.
Thus, the helicity-1 and helicity-2 equations have well-defined Cauchy problems.

The problem lies entirely with the helicity-0  equation which can be rearranged to take the form
\ba
\mathcal{E}_{\pi}+ \frac{1}{3 \Lambda_5^5}\partial^{\omega} \partial^{\nu} \left[ \partial^{\mu} \partial_{\omega} \pi \,  \mathcal{E}_{h_{\mu \nu}} \right] = \Box \pi +\frac{1}{m^2} \partial_{\mu} \partial_{\nu}S^{\mu\nu} + \frac{1}{2} S +\frac{1}{6 \Lambda_5^5}\p^{\omega} \p_\nu \left[ \p_\mu \p_{\omega} \pi \, S^{\mu\nu} \right] =0\,.
\label{eqpi}
\ea
If no conditions are placed on the external source, this equation includes higher derivatives of $\pi$ at the scale $\Lambda_5$ so the ghost or perturbation theory problems would seem to be reintroduced. However, in any realistic case the matter source itself will depend on the spin-2 field. Is it possible to couple the entire spin-2 field $h_{\mu\nu}$ to matter in such away that perturbation theory is well-defined at the scale $\Lambda_5$ (and even up to the scale $\Lambda_3$) and there are no ghosts? The clue to the resolution is that the source in \eqref{vector_mat} appears to diverge as $m \rightarrow 0$, \ie in the decoupling limit that has been taken. Thus in order to have a consistent decoupling limit it is necessary for the divergence of the source to behave as $\partial^{\mu} S_{\mu\nu} \sim m^2 \rightarrow 0$. The fact that at leading order the stress energy must be conserved is a reflection of the fact that in the $m \rightarrow 0$ limit, the helicity-2 mode exhibits a linearized diffeomorphism symmetry $\tilde{h}_{\mu\nu} \rightarrow \tilde{h}_{\mu\nu} + \partial_{\mu} \chi_{\nu}+ \partial_{\nu} \chi_{\mu}$ such that the helicity-1 and helicity-0 modes are invariant (since their variation is of order $m$). This symmetry that arises in this limit then forces that its source is conserved (at leading order) just as in GR and so  $\partial^\mu S^{(0)}_{\mu\nu} =0$. Of course, this symmetry is not exact, and so the source fails to be conserved at order $m^2$ such that $ \partial^{\mu}S_{\mu\nu}/m^2$ is finite in the limit $m \rightarrow 0$. This ensures that the sources in equations \eqref{vector_mat} and \eqref{tildeh_mat} are finite. In this way we can see that there is a well-defined decoupling limit at the scale $\Lambda_5$ even in the presence of a source.
The implication is that the helicity-1 and helicity-0 modes are sourced by terms which are subleading in the decoupling limit from the point of view of the matter fields. These subleading terms account for the non-conservation of the source due its backreaction, \ie the backreaction of gravity, on the equations of motion for the matter field. However this in turn implies that 
$\partial^{\mu}S_{\mu\nu}/m^2$ must itself depend on the spin-2 field. The question we must address then is: Is it possible in a realistic matter system to couple to matter in such a way that the failure of conservation of the source because of the gravitational backreaction onto the matter implies that the net source in \eqref{eqpi} does not dependent on higher derivatives of the helicity-0 mode?

\subsubsection{Consistency Relation for matter}

In practice this is all extremely easy to achieve. For instance, suppose the matter couples to the spin-2 field in a way that preserves diffeomorphism invariance. This is a choice that can always be made classically. $S^{\mu\nu}$ then takes the same expression it would do in GR, \ie $S^{\mu\nu}$ is just proportional to the usual stress energy of the matter field $S^{\mu\nu}= \sqrt{-g} \, T^{\mu\nu}/\mpl $. However, as is well known, this stress energy is not ordinarily conserved but is rather covariantly conserved as a direct consequence of the backreaction of gravity on the matter:
\ba
\sqrt{-g}\, D_\mu T^{\mu\nu}=\p_\mu(\sqrt{-g}\, T^{\mu\nu})+ \sqrt{-g} \Gamma^\nu_{\mu \omega}\, T^{\mu\omega}=0\,,
\ea
so that we can write the covariant conservation law for $S^{\mu\nu}$ as
\ba
\p_\mu \( S^{\mu\nu}\) = - \Gamma^\nu_{\mu \omega} \, S^{\mu \omega}\,.
\ea
Here the Christoffel symbol should be evaluated on the full metric, however in the context of the $\Lambda_5$ decoupling limit it is sufficient to evaluate it with only the leading part of the helicity zero mode contribution to the metric included $\delta g_{\mu\nu} = \frac{1}{3\mpl  m^2} \partial_{\mu} \partial_{\nu} \pi $ since this is the term that dominates in the $\Lambda_5$ decoupling limit. This gives
\ba
 \Gamma^{\nu}_{\mu \omega}= \frac{1}{6 \mpl  m^2} \p^\nu \p_\omega \p_\mu \pi + \dots\,.
\ea
Thus to leading order in the decoupling limit we have
\ba
\frac{1}{m^2} \p_{\mu} \p_{\nu}  S^{\mu\nu} &=& - \frac{1}{m^2} \p_\omega \(  \Gamma^{\omega}\mn \, S^{\mu \nu} \) \\
  &=&-\frac{1}{6 \Lambda_5^5} \p_\omega \( \p^\omega \p_\mu \p_\nu \pi S^{\mu \nu} \) + \dots \\
  &=& -\frac{1}{6 \Lambda_5^5}\,  \p_\omega \p_\nu \(\p^\omega \p_\mu  \pi S^{\mu \nu} \)+\mathcal{O}(\Lambda_4^{-4})+\dots
\ea
where in the last step we made use of the fact that the failure of conservation of $S^{\mu\nu}$ is a term only relevant at a higher scale. Putting this together, at the scale $\Lambda_5$ the correct equation of motion for the helicity-0 mode, in the decoupling limit $m \rightarrow 0 $ keeping $\Lambda_5$ fixed is the perfectly well-defined equation
\ba
\mathcal{E}_{\pi}+ \frac{1}{3 \Lambda_5^5}\p^\omega \p^\nu \left[ \p_\mu \p_\omega \pi \,  \mathcal{E}_{h\mn} \right] = \Box \pi + \frac{1}{2} S =0\,.
\ea
Thus even in the presence of matter the equations of motion remain second order, and the total number of propagating modes in the gravity sector remains as 5. Furthermore as with previous arguments it is trivial to see that perturbation theory remains well behaved at the scale $\Lambda_5$ since it terminates at first order, as in the scalar toy model.

\subsubsection{Preserving approximate linearized diffeomorphism invariance}

We made this argument assuming that matter couples in a way that preserves diffeomorphism invariance. Since diffeomorphism invariance is broken in massive gravity we can in principle imagine other couplings to matter. However it is easy to see that regardless of the coupling to matter it is never possible to reintroduce problems at the scale $\Lambda_5$. The reason is that as we have already explained, in the limit $m \rightarrow 0$ the source $S_{\mu\nu}$ must be conserved $\partial^{\mu} S_{\mu\nu} \rightarrow 0$. Thus at leading order this source must be the stress energy of the matter field. As in GR, adding the term $\frac{1}{2 \mpl }h_{\mu\nu} T^{\mu\nu}$ back into the Lagrangian then causes the matter fields to be sourced by its own stress energy in such away that the stress energy fails to be conserved at order $1/\mpl $. However, the way it fails to be conserved to this order is just the same as in GR and so the above argument applies. In other words, any couplings to matter which do not respect diffeomorphism invariance would have to come in at order $h_{\mu\nu}^2$ or higher, and so would not contribute in the $\Lambda_5$ decoupling limit. In fact this argument extends up to the scale $\Lambda_3$ since this is the scale at which $h_{\mu\nu} \sim 1$ and perturbation theory breaks down. \\

In conclusion, it is not possible to reintroduce any problem at scales below $\Lambda_3$ in coupling to matter, the approximate symmetry $\tilde{h}_{\mu\nu} \rightarrow \tilde{h}_{\mu\nu} + \partial_{\mu} \chi_{\nu}+ \partial_{\nu} \chi_{\mu}$ is sufficient to protect the couplings to matter even at the quantum level in such a way that below the scale $\Lambda_3$ the equation of motion for the helicity modes are all well-defined, propagating no more that 5 degrees of freedom, and perturbation theory remains well behaved in the presence of matter. Notice that at the scale $\Lambda_3$ the resulting theory of massive gravity is  similar to that of DGP, suggesting that the treatment of both theories could be thought in similar terms.

\subsubsection{A specific example of consistent coupling to matter}

One simple example of how one can consistently couple to matter working in the helicity variables is provided by a point particle of mass $m_p$ described by the trajectory $X^{\mu}(\tau)$ with action
\be
S_{\rm matter}  = -m_p \int \d \tau \( -\frac{1}{2e} g_{\mu\nu} \frac{\d X^{\mu}}{\d \tau}  \frac{\d X^{\nu}}{d \tau} + \frac{1}{2}e \)
\ee
where $\tau$ denotes the proper time along the trajectory, and $e$ is the einbein which accounts for the worldline reparameterization invariance. To have a consistent $\Lambda_5$ decoupling limit we scale $m_p \rightarrow \infty$ such that $m_p/\mpl$ is finite. In this case $S^{\mu\nu}$ is given by
\be
S^{\mu\nu} = \frac{m_p}{M_{\rm Pl}} \int \d \tau \, \delta^4(x-X) \frac{1}{e} \frac{\d X^{\mu}}{\d \tau}  \frac{\d X^{\nu}}{\d \tau}.
\ee
By the previous arguments, one can show that using the equations of motion for the particle, ensures that the coupling to $\pi$ in the $\Lambda_5$ decoupling limit is carried entirely by $S$. However, we can also see this more directly by performing a field redefinition of the matter variables. Working with the full action and making the helicity decomposition we have
\ba
&& \nn g_{\mu\nu} \frac{\d X^{\mu}}{\d \tau}  \frac{\d X^{\nu}}{d \tau}
 = \left( \eta_{\mu\nu} + \frac{\hat{h}_{\mu\nu}}{\mpl}+\frac{1}{2m \mpl} \partial_{( \mu} A_{\nu)}+\frac{1}{3m^2\mpl} \partial_{\mu}\partial_{\nu} \pi + \frac{1}{6\mpl} \pi \eta_{\mu\nu}\right) \frac{\d X^{\mu}}{d \tau}  \frac{\d X^{\nu}}{d \tau}  \\ \nn
&&  = \left( \eta_{\mu\nu} + \frac{\bar{h}\mn}{\mpl}+\frac{1}{36\Lambda_5^5} \Pi^2_{\mu \nu}+\frac{1}{2m \mpl} \partial_{( \mu} A_{\nu)}+\frac{1}{3m^2\mpl} \partial_{\mu}\partial_{\nu} \pi + 
 \frac{1}{6\mpl} \pi \eta_{\mu\nu}\right) \frac{\d X^{\mu}}{d \tau}  \frac{\d X^{\nu}}{\d \tau}\,.
\ea
Let us focus on one set of terms in this expression. After integration by parts we see that
\ba
\nn
\int \d \tau \frac 1 e\( \frac{1}{2m \mpl} \p_{( \mu} A_{\nu)}+\frac{1}{3m^2\mpl} \p_{\mu}\partial_{\nu} \pi  \)\frac{\d X^{\mu}}{\d \tau}  \frac{\d X^{\nu}}{\d \tau} \hspace{40pt} \\ 
= \int \d \tau \frac{\d}{e\, \d \tau}\( \frac{1}{m\mpl} A_{\nu}+ \frac{1}{3m^2\mpl} \p_{\nu}\pi\)\frac{\d X^{\nu}}{\d\tau}\,.
\ea
Similarly
\ba
\nn
\int \d \tau \frac 1 e \frac{1}{36\Lambda_5^5} \Pi^2\mn \frac{\d X^{\mu}}{d \tau}  \frac{\d X^{\nu}}{\d \tau} = \int \frac{\d \tau} {e \, 36\Lambda_5^5} \frac{\d \p_\mu \pi}{\d \tau}\frac{\d \p^\mu \pi}{\d \tau}\,.
\ea
Now let us perform the following well-defined, invertible field definition
\ba
Y^{\mu} = X^{\mu}+\frac{m^3}{2 \Lambda_5^5}A^{\mu}+\frac{m^2}{6\Lambda_5^5} \p^\mu \pi\,.
\ea
It is easy to see that up to terms which vanish in the $\Lambda_5$ decoupling limit the matter action becomes
\be
S_{\rm matter}  = -m_p \int \d \tau \( -\frac1{2e} \( \eta\mn + \frac{\bar{h}\mn}{\mpl}+\frac{1}{6\mpl} \pi \eta\mn\) \frac{\d Y^{\mu}}{\d \tau}
\frac{\d Y^{\nu}}{\d \tau} + \frac{1}{2}e \) + {\cal O}(1/\Lambda_4^4)\,.
\ee
Since as we have already discussed the kinetic terms of $\bar{h}\mn$ and $\pi$ decouple at this order, it is transparent from the above action that that $\pi$ only couples to $S$ and that the equations of motion are second order. The field redefinition of the matter variables remains under perturbative control since $Y^\mu-X^\mu$ vanishes as $m^2$ in the $\Lambda_5$ decoupling limit. As usual, we did not need to perform the field redefinition to see that the equations of motion of the spin-two field and the particle were both well-defined, however it certainly simplifies the analysis.

As discussed earlier, the above result would be identical had we modified the action for the particle by adding terms which violate full diffeomorphism invariance. For example, if we add terms of the form
\be
\Delta S_{\rm matter} = -m_p \int \d \tau \frac{1}{e} F^{\mu\nu \alpha\beta}\left(X^{\omega},\frac{1}{e}\frac{dX^{\omega}}{d\tau}\right) \frac{h_{\mu\nu}}{\mpl}\frac{h_{\alpha\beta}}{\mpl} + \dots
\ee
where $F^{\mu\nu \alpha\beta}$ is a tensor constructed locally out of the particles position and velocity. This choice is a natural one since we still require the equations of motion for the particle to remain second order. This term clearly violates full diffeomorphism invariance, nevertheless it is harmless at this order because it vanishes in the $\Lambda_5$ (and $\Lambda_4$) decoupling limit with the natural assumption that $F$, being dimensionless, is kept fixed in this limit. Similar reasoning applies to terms with more powers of the spin-2 field, or more derivatives. Thus the fact that in massive gravity the coupling to matter is less constrained is not in itself necessarily a problem.

To reiterate, the linearized diffeomorphism invariance which always arises accidently in the decoupling limit, forces the leading order coupling to matter $\frac{1}{2 \mpl}h_{\mu\nu}T^{\mu\nu}$ to be the same as in GR, and is sufficient to guarantee the absence of ghosts/breakdown of perturbation theory below the scale $\Lambda_3$ at least. At the scale $\Lambda_3$ the situation becomes more subtle because at this scale the helicity-0 mode becomes strongly coupled. However, even at this scale, whilst loops from the helicity-0 mode will in principle generate matter couplings which do not respect full diffeomorphism invariance, these will inevitably be suppressed by additional powers of $M_{\rm pl}$ and thus occur at a scale higher that $\Lambda_3$. Since the helicity decomposition is a bad one in this regime, it is beyond the scope of this work to understand the full quantum consistency of couplings to matter in massive gravity.

\subsection{Consistency of perturbation theory in the $\Lambda_5$ theory}

Finally let us come to the issue of  the scale at which perturbation theory breaks down. For scattering processes we have seen that the theory is free at the scale $\Lambda_5$ and so perturbation theory is valid (as we will see below up to the scale $\Lambda_3$). However, let us consider what happens for spherically symmetric configurations in the presence of a source of mass $M$, localized at $r=0$. Following standard reasoning in the weak field region the solutions for $\tilde{h}$ and $\pi$ take the form $\sim \frac{M}{\mpl  r}$. In the original helicity variables, the $\Lambda_5$ terms do have an important role to play in the classical solutions since at some characteristic distance, the term $(\partial \partial \pi)^2/\Lambda_5^{5} \sim \frac{M^2}{\mpl ^2\Lambda_5^{5} r^6 }$ becomes comparable to $\tilde{h} \sim  \frac{M}{\mpl  r}$. Thus in these variables the apparent interactions seem to become important at a scale $r \sim (M \Lambda_3^{-6})^{1/5}$. However this does not signal the breakdown of perturbation theory for the spin-2 field. As in the two scalar toy-model, the perturbative expansion actually terminates at first order in perturbations in the $\Lambda_5$ decoupling limit. Thus {\bf not only is perturbation theory well behaved, it is exact at first order! }\\

This is all made clear by actually solving the equations of motion in the presence of matter, which by the above arguments are seen to be of the form
\ba
&&\mathcal{E}_{\pi}+ \frac{1}{3 \Lambda_5^5}\partial^{\omega} \partial^{\nu} \left[ \partial_{\mu} \partial_{\omega} \pi \,  \mathcal{E}_{h\mn} \right] = \Box \pi + \frac{1}{2\mpl } T =0 \label{eom1} \\ \label{eom2}
&& \mathcal{E}_{A_{\nu}}=\partial^{\mu} F_{\mu\nu}=0 \, , \\
&&\mathcal{E}_{h\mn}={\Ein}^{\alpha \beta}\mn \(\tilde{h}_{\alpha \beta} - \frac{1}{36\Lambda_5^{5} }\Pi^2_{\alpha\beta}\)+\frac{1}{2\mpl  }T\mn=0\,.
\ea
The equation \eqref{eom1} may be solved exactly without any perturbative corrections
\ba
\pi = \pi^0 - \frac{1}{2\mpl}\frac 1 \Box\, T\,,
\ea
where $\pi^0$ as a solution of the homogeneous equation $\Box \pi_0=0$.
The vector equation may be solved similarly, {\it e.g.} by choosing Lorentz gauge $\partial_{\mu}A^{\mu}=0$
\ba
A_\nu = A_\nu^0 \,,
\ea
for which $\Box A_\nu^0=0$ and finally by choosing de Donder gauge $\partial^\mu(\tilde h\mn- \frac{1}{2} \eta\mn \tilde h)=0$ and substituting in the solution for $\pi$ we obtain
\ba
\tilde h\mn = \tilde h\mn^{0} -\frac{1}{2\mpl}\frac 1 \Box \( T\mn- \frac{1}{2} \eta\mn T\) + \frac{1}{18 \Lambda_5^5}\frac 1 \Box V\mn|_{\pi=\pi^0 - \frac{1}{2\mpl}\frac 1 \Box\, T}  \,.
\ea
This completes the exact solutions for an arbitrary source. These solutions are also the ones that arise at first order in perturbation theory since the contribution to $\tilde h_{\mu\nu} $ is linear in $1/\Lambda_5^5$. This we see that indeed first order perturbation theory is exact, and so perturbation theory is certainly under control. \\

Another way of understanding this result is when the $\Lambda_5$ terms become important, the dimensionless spin-2 field $h_{\mu\nu}/\mpl  \sim (m r_s)^{4/5} \ll 1$ where $r_s$ is the Schwarzschild radius of the source. Since for realistic values this is negligible, we see that perturbation theory of the spin-2 field is easily under control at this scale. Pursuing this argument to its end, we recover the known result that the actual scale of the breakdown of perturbation theory, \ie the scale at which the perturbative expansion fails to converge, is the scale $\Lambda_3$, the correct strong coupling scale of the helicity-0 mode, since it is only at this scale that $h_{\mu\nu} \sim 1$.

\section{Relevant interactions at the scale $\Lambda_4$}
\label{sec:L4}
Now that we have checked the consistency of the theory at the scale $\Lambda_5$, let us go beyond that scale and look for the relevant interactions all the way up to the scale $\Lambda_4$.

There are potentially dangerous interactions that arise at this order from mixing the vectors and scalars. For instance in \cite{Folkerts:2011ev} it was pointed out we could get already at cubic order terms of the form $(\p A) (\p^2 \pi)^2$. Besides these, there is an entire zoo of interactions entering at the scale $\Lambda_4$. As in the previous section, it is essential to keep track of every single one of them before determining the consistency of the theory, since all the Feynman diagrams at the same scale contribute to the amplitude of scattering processes. Many of them taken alone would break perturbative unitarity, but as we shall see below, for the specific mass term considered in \eqref{Lm}, the sum of all these diagrams does not violate unitarity and leads to a trivial S-matrix at that scale.

The interactions entering at the scale $\Lambda_4$  are as follows:
\begin{enumerate}
\item At the cubic level, $\p A (\p^2 \pi)^2$ from the mass term and $\p^2 \tilde h \p A \p^2 \pi$,  $\p A (\p^2 \pi)^2$  and $ \p^2 (\p A)^3$ from the EH term.
\item At the quartic level, $(\p^2 \pi)^4$ from the mass term and $\p^2 \tilde h (\p^2 \pi)^3$, $(\p^2 \pi)^4$ or $(\p A)^2 \p^2 (\p^2 \pi)^2$ from the EH term.
\item At the quintic level, $(\p A)  \p^2 (\p^2 \pi)^4$ from the EH term.
\item And finally, at the sextic level, $(\p^2 \pi)^3 \p^2 (\p^2 \pi)^3$ from the EH term.
\end{enumerate}
Furthermore, there are also two interactions in the intermediate region between  $\Lambda_5$ and $\Lambda_4$, the first one occurs at quartic order and is of the form $(\p A)\p^2 (\p^2 \pi)^3$ coming at the scale $\Lambda_{9/2}$ and the second one occurs at quintic order and is of the form $\p^2 (\p^2 \pi)^5$ with the slightly higher scale $\Lambda_{13/3}$.

With the correct choice of mass \eqref{Lm}, many of these terms combine to give rise to a total derivative.
The rest of them can be treated in a similar manner to the terms that arise at the scale $\Lambda_5$.
One can easily see that up to total derivatives, one has
\ba
\mathcal{L}^{(3)}_{\Lambda_4,\ \p^2 (\p A)^3}=0\,.
\ea
As for the other interactions, we can start by focusing on terms of the form $(\p A) (\p^2 \pi)^2$ at the cubic order.

\subsection{$A \to \pi \pi$ scattering Amplitude}

Although on-shell the 3-point function $\langle A \pi \pi \rangle$ vanishes in the decoupling limit by energy/momentum conservation, it does not necessarily do so off-shell, {\it e.g.} when looking at perturbations around a given background. Since its off-shell behaviour is important in higher order diagrams let us consider the off-shell form of its amplitude. This has a contribution from the following interaction
\ba
\mathcal{L}^{(3)}_{\Lambda_4,\ (\p A) (\p^2 \pi)^2}=\frac 1{36\Lambda_4^4}A^\mu\Big[
s_1 \p_\mu [\Pi]^2+s_2 \p_\mu [\Pi^2]
+s_3 \p_\alpha \Box \pi\p_\mu \p^\alpha \pi \Big]\,,
\ea
with $s_1=-(1+4k_2+12k_3)$, $s_2=(1-6k_1-4k_2)$ and $s_3=-(6k_1+4k_2)$. With the coefficient  $k$'s given in \eqref{k's} this simplifies to
\ba
\label{L3Apipi}
\mathcal{L}^{(3)}_{\Lambda_4,\ (\p A) (\p^2 \pi)^2}=\frac 1{36\Lambda_4^4}A^\mu\Big[
\p_\mu [\Pi^2]- \p^\nu \Pi^2\mn \Big]\,.
\ea
The existence of this interaction, taken alone, would imply the breakdown of perturbative unitarity at the scale $\Lambda_4$. Furthermore, since this interaction involves three derivatives on $\pi$, the 3-point scattering process $ A \to \pi \pi $ would appear to contain a ghost. If this was the end product, this scattering process would not have a well-posed Cauchy problem and the theory would have a ghost. Fortunately, there is another class of diagrams that contributes to this amplitude which arises from the tree level process $A \to \tilde h \to \pi \pi$. \\

The first decay $A\to \tilde h$ is governed by the interaction  $-m A_\mu \(\p_\nu \tilde h^{\mu\nu}-\p_\mu \tilde h\)$ in \eqref{L2tot}. A priori this term is negligible as it suppressed by a factor $m/k$. However the second decay $\tilde h \to \pi \pi$ is governed by the interaction  $-\frac{1}{36 \Lambda_5^5}\tilde h^{\mu\nu}\Ein^{\alpha\beta}\mn \Pi^2_{\alpha\beta}$ found in \eqref{L3_5} and is enhanced by a factor  $\Lambda_5^5/k^5\sim k/m$. While the first decay, is very unlikely to occur at an energy scale $\Lambda_4$ or below, the second decay on the other hand is extremely fast, such that this channel is actually as efficient as the direct channel $ A \to \pi \pi $. Combining these two diagrams together, it is easy to see that the resulting amplitude of the 3-point function $\langle A \pi \pi \rangle$ vanishes at tree level at the scale $\Lambda_4$ or below.
\vspace{20pt}

\begin{figure}[!htb]
\begin{center}
\begin{fmffile}{APiPi}
\parbox{33mm}{\begin{fmfgraph*}(80,40)
	            \fmfleft{i1}
	            \fmfright{o1,o2}
                \fmflabel{$A_\mu$}{i1} 
                \fmflabel{$\pi_1$}{o1} 
                \fmflabel{$\pi_2$}{o2} 
                \fmflabel{$\phantom{.}\  \frac {\p^5}{36\Lambda_4^4}$}{v1} 
	            \fmf{wiggly}{i1,v1}
                \fmf{plain}{v1,o1}
                \fmfdot{v1}
                \fmf{plain}{v1,o2}
\end{fmfgraph*}} + \hspace{30pt}
\parbox{40mm}{\begin{fmfgraph*}(120,40)
	            \fmfleft{i1}
	            \fmfright{o1,o2}
                \fmflabel{$A_\mu$}{i1} 
                \fmflabel{$\pi_1$}{o1} 
                \fmflabel{$\pi_2$}{o2} 
	            \fmf{wiggly}{i1,v1}
                \fmf{dashes,label=$\tilde h\mn$}{v1,v2}
                \fmf{plain}{v2,o1}
                \fmfdot{v1,v2}
                \fmf{plain}{v2,o2}
                \fmflabel{$\vspace{20pt}  { m \p \hspace{-17pt} \phantom{.}} $}{v1}
                \fmflabel{$-\frac {\p^6}{36\Lambda_5^5}$}{v2}
\end{fmfgraph*}} \hspace{20pt} = \hspace{5pt} 0
\end{fmffile}
\end{center}
\label{fig2}
\caption{Tree-level diagrams contributing to the 3-point function $\langle A \pi \pi \rangle$ at the scale $\Lambda_4$.}
\end{figure}
Of course the fact that we must work hard to see this cancelation is just a reflection of the fact that to look at physics at the scale $\Lambda_4$, it is better to work with the correctly diagonalized helicity-2 mode defined by the scale $\Lambda_5$. Had we done this from the outset neither of these processes would have arisen in the first place.

\subsection{$\pi \pi \to \pi \pi$ scattering Amplitude}
Another type of interaction that arises on-shell and deserves attention is the quartic one of the form $(\p^2 \pi)^4$, which gives a contribution
\ba
\label{pi4_Lambda_4}
\mathcal{L}^{(4)}_{\Lambda_4,\ (\p^2 \pi)^4}&=&-\frac{1}{2\times36^2\Lambda_4^8}\([\Pi^4]-[\Pi^2]^2\)\,.
\ea
Here again, this interaction has non-vanishing higher derivative terms which would seem to suggest a problem. However, at the same scale the channel
$\pi \pi \to \tilde h \to \tilde h \to \pi \pi$ gives precisely an opposite contribution which cancels \eqref{pi4_Lambda_4}. This channel can be understood as follows: First both $\pi$'s decay into a helicity-2 mode via the interaction  $-\frac{1}{36 \Lambda_5^5}\tilde h^{\mu\nu}\Ein^{\alpha\beta}\mn \Pi^2_{\alpha\beta}$ found in \eqref{L3_5} at the scale $\Lambda_5$. Then the interaction $-\frac 12 m^2\(\tilde h\mn^2-\tilde h^2\)$ in \eqref{L2tot} is suppressed by $m^2$ (strictly speaking this should be part of the helicity-2 propagator, but since this has been neglected for the previous interactions, this has to be included as an ``additional" interaction at this level). Finally, the helicity-2 decays back into 2 helicity-0 via the inverse process $\tilde h \to \pi \pi$ which occurs at the scale $\Lambda_5$. Putting all this together, this gives rise to three new diagrams with an effective scale $(\Lambda_5^{10}/m^2)^{1/8}=\Lambda_4^4$ which should therefore be considered at the same time as \eqref{pi4_Lambda_4}. We can easily see that the sum of all these four diagrams precisely cancels. \\

As we can see, this is a very cumbersome approach to compute these scattering processes, and the most direct way to deal with this is simply to diagonalize the helicity-2 mode at the cubic level through the redefinition \eqref{redef1}. We emphasize however once again, that whilst it makes more sense to work in terms of the diagonalized field, no physics is hidden in this redefinition, and we have gone to great length to show explicitly how to understand  the physics if we were to chose to work in terms of the undiagonalized field. We now turn back to a more conventional approach and work in terms of the $\bar h\mn$ defined in   \eqref{redef1}.

\subsection{Free theory at the scale $\Lambda_4$}
Except for the three interactions $\p^2 \tilde h \p A \p^2 \pi$,  $(\p A^2)^2\p^2 (\p^2 \pi)^2$ and $(\p A)\p^2 (\p^2 \pi)^3$ all the other interactions arising at or below the scale $\Lambda_4$ are nothing else but part of the diagonalized helicity-2 $\bar h\mn$ interactions. More precisely,
\ba
&& \mathcal{L}^{(3)}_{\Lambda_4,\ (\p A) (\p^2 \pi)^2} \subset -m A_\mu \(\p_\nu \bar h^{\mu\nu}-\p_\mu \bar h\)\\
&& \mathcal{L}^{(4)}_{\Lambda_4,\ (\p^2 \pi)^2}\subset-\frac 12 m^2\(\bar h\mn^2-\bar h^2\)\\
&& \mathcal{L}^{(4)}_{\Lambda_4,\ (\p^2 \tilde h)(\p^2 \pi)^2}\subset\frac{1}{\mpl m^2} \bar h\p^2 \bar h (\p^2 \pi)\\
&& \mathcal{L}^{(5)}_{\Lambda_4,\ (\p A) (\p^2 \pi)^4}\subset\frac{1}{\mpl m} \p A \bar h \p^2 \bar h\\
&& \mathcal{L}^{(6)}_{\Lambda_4,\ (\p^2 \pi)^3 \p^2 (\p^2 \pi)^3}\subset\frac{1}{\mpl} \bar h^2 \p^2 \bar h\\
&& \mathcal{L}^{(5)}_{\Lambda_{13/3},\ \p^2 (\p^2 \pi)^5}\subset\frac{1}{\mpl m^2} \bar h\p^2 \bar h (\p^2 \pi)\,,
\ea
so after appropriate diagonalization of the helicity-2, the previous terms do not appear in the theory. In other words, if we were to keep working in terms of $\tilde h\mn$, any diagram involving the previous vertices precisely cancels with another diagram such that the resulting scattering amplitude in question vanishes.
Once we work with $\bar h\mn$, the only three remaining terms that arise between the scale $\Lambda_5$ and $\Lambda_4$ included are given by,
\ba
&& \mathcal{L}^{(3)}_{\Lambda_4,\ \p^2 \bar h \p A \p^2 \pi}=-\frac{1}{12\Lambda_4^4}\bar h^{\mu\nu}\
\hat{\mathcal{E}}^{\alpha\beta}\mn\ P_{\alpha\beta}\\
&& \mathcal{L}^{(4)}_{\Lambda_4,\ (\p A)^2\p^2 (\p^2 \pi)^2}=\frac{1}{8\Lambda_4^8} P^{\mu\nu}\
\hat{\mathcal{E}}^{\alpha\beta}\mn\ P_{\alpha\beta}\\
&& \mathcal{L}^{(4)}_{\Lambda_{9/2},\ (\p A)\p^2 (\p^2 \pi)^3}=\frac{1}{12 \times 36 \Lambda_4^4 \Lambda_5^5} P^{\mu\nu}\
\hat{\mathcal{E}}^{\alpha\beta}\mn\ \Pi^2_{\alpha\beta}\,,
\ea
with $P\mn=\p_{(\mu}A^\alpha\Pi_{\nu)\alpha}$.
All these interactions therefore combine to form
\ba
\mathcal{L} &=&-\frac 18 F\mn^2 +\frac 1{12} \pi \Box \pi\nn\\
&+&\frac12 \(\bar h^{\mu\nu}-\frac{P^{\mu\nu}}{12 \Lambda_4^4}\) \Ein^{\alpha \beta}\mn
 \(\bar h_{\alpha\beta}-\frac{P_{\alpha \beta}}{12 \Lambda_4^4}\) \label{dec4} +\cdots\,,
\ea
where the dots denote terms that vanish in the decoupling limit if $\bar h$ is kept finite.
It is easy to see that the second line does lead to higher derivative interactions terms, which by themselves break unitarity, but the combined sum of the diagrams from interactions on the second line is of course fine as has already been discussed at length previously. The remaining coupling between $\bar h\mn$ and $P\mn$ at quartic order, is nothing else but the indication that $\bar h$ has still not been fully diagonalized, and the properly diagonalized helicity-2 mode (up to the scale $\Lambda_4$) is instead
\ba
\label{brevered}
\breve{h}\mn=\tilde h\mn-\frac{1}{36\Lambda_5^{5}}\Pi^2\mn-\frac{1}{12 \Lambda_4^4}P\mn\,.
\ea
Working in terms of the helicity-2 mode $\breve{h}$, which is the one whose kinetic terms is diagonalized around an arbitrary background at least up to the scale $\Lambda_4$, the decomposition of field reads as follows:
\ba
\label{gettingThere}
h\mn=\breve{h}\mn+\frac 16 \pi \eta\mn+\mpl (\Psi_{\mu\alpha}\Psi_\nu^{\ \alpha}-\eta\mn)
-\frac{1}{4\Lambda_3^3}\p_{\mu}A_\alpha \p_\nu A^\alpha\,,
\ea
with
\ba
 \Psi_{\mu\alpha}=\eta_{\mu\alpha}+\frac{1}{2\mpl m}\p_\mu A_\alpha+\frac{1}{6\mpl m^2}\p_{\mu}\p_{\alpha}\pi\,.
\ea
and the Lagrangian can again be seen to be that of a free theory
\ba
\mathcal{L} =-\frac 18 F\mn^2 +\frac 1{12} \pi \Box \pi+\frac12 \breve h^{\mu\nu} \Ein^{\alpha \beta}\mn
 \breve h_{\alpha\beta} \label{dec41} +\mathcal{O}(\Lambda^{-n}_{n})\,,
\ea
where corrections come in at a larger energy scale $n<4$.

\subsection{Validity of perturbation theory and coupling to matter}

We may now simply apply all the arguments used at the scale $\Lambda_5$ to see that perturbation theory is under control at the scale $\Lambda_4$. As before the point is that the $\pi$ equation of motion can be seen to be
\ba
\Box \pi=0
\ea
in the absence of a source. As before, taking into account the conditions that must be satisfied by any allowed coupling to matter we infer that in the presence of a source the equations also remain second order
\ba
\Box \pi = -\frac{1}{2\mpl } T.
\ea
This means it is always possible to determine $\pi$ to all orders in perturbation theory. Given this solution we may then solve the well-defined second order equations for the helicity-2 and helicity-1 modes and as before perturbation theory is seen to be exact to first order in $1/\Lambda_5^5$ and $1/\Lambda_4^4$ something which is made manifest by the field redefinition \eqref{brevered}.

\section{Beyond $\Lambda_4$}
\label{sec:L3}

\subsection{Between $\Lambda_4$ and $\Lambda_3$}
Having shown that no pathological interactions arise at energy scales $k$ lower or equal to $\Lambda_4$, let us consider the region  $\Lambda_4<k<\Lambda_3$. Since the last term in \eqref{gettingThere} is negligible at these scales, we may simply omit it and check the consistency of the theory without it. As a simple consequence of local diffeomorphism invariance, the EH term can be computed without the contribution from $\Psi$,
 \ba
 &&\mpl ^2\sqrt{-g}R|_{g\mn=\frac{1}{\mpl}(\breve{h}\mn+\frac 1{6} \pi \eta\mn) +\Psi_{\mu\alpha}\Psi_\nu^{\ \alpha}} =\\\ \nn
 && \quad \mpl ^2 \sqrt{-g}R|_{g\mn=\eta\mn + \frac{1}{\mpl}(\breve{h}\mn+\frac 1{6} \pi \eta\mn)} + \text{terms which vanish below $\Lambda_3$}\,,
 \ea
so between the scale $\Lambda_4$ and $\Lambda_3$ there cannot be any additional contribution from the EH term. Any contribution should therefore come from the mass term. Fortunately only two classes of interactions are relevant at these scales, namely terms of form $(\p^2 \pi)^n$ arising at the scale $\Lambda_{(3n-4)/(n-2)}$ (with $n>4$), and the ones of the form $(\p A)(\p^2 \pi)^n$ arising at the scale $\Lambda_{(3n-2)/(n-1)}$ (with $n>3$). The mass term we have considered in \eqref{Lm} has been specifically engineered so as to cancel all the interactions of the form   $(\p^2 \pi)^n$
and 
we will not reproduce this calculation here. The second type of interactions,  $(\p A)(\p^2 \pi)^n$ on the other hand deserves special care. Upon close inspection, one can check that all the terms of that form resum to form the following all-orders Lagrangian
\ba
\label{dA X}
\mathcal{L}_{(\p A)(\p^2 \pi)^n}=\mpl m \(\p^\mu A^\nu +\p^\nu A^\mu\)X\mn\,,
\ea
with
\ba
X\mn&=&\frac{1}{\mpl m^2}\(\Pi \eta \mn -\Pi\mn\)
+\frac{1+3\alpha_3}{\mpl^2 m^6}\(\Pi\mn^2-\Pi\, \Pi\mn+\frac 12 \([\Pi]^2-[\Pi^2]\)\eta\mn\) \\
&+&\frac{\alpha_3+4\alpha_4}{\mpl^3 m^9}\Bigg(\Pi\mn^3-\Pi\, \Pi\mn^2+\frac 12 \Pi\mn \([\Pi]^2-[\Pi^2]\)
-\frac 16 ([\Pi]^3-3[\Pi][\Pi^2]+2[\Pi^3])\eta\mn\Bigg)\,.\nn
\ea
Since $X\mn$ is identically transverse $\partial^{\mu} X_{\mu\nu}=0$, the final interaction term \eqref{dA X} that could survive below $\Lambda_3$ is actually a total derivative and gives no contribution
(in any case, we see that this would never give anything beyond $n=3$).
The theory at hand is therefore free all the way up to (but not including) the scale $\Lambda_3$. In other words we have finally shown that the correct scale of interactions of the helicity modes is the scale $\Lambda_3$. Thus the hierarchy of scales that seems to arise between $\Lambda_5$ and $\Lambda_3$ is entirely fake, no physics actually occurs at these scales, no interactions occur at these scales, no ghosts appear at these scales, perturbation theory of the spin-2 field remains under control at these scales, and last but not least, there are no quantum corrections at these scales !

\subsection{Relevant interactions at the scale $\Lambda_3$: The true strong coupling scale}

Finally we come to the scale $\Lambda_3$. At this scale the helicity-0 mode has genuine interactions, and it is easy to show that the helicity-0 becomes strongly coupled at this scale. Tree level unitarity is violated at this scale. This is not a new statement, but it follows equally in the helicity decomposition. When this occurs the spin-2 perturbation theory breaks down as well, and so this is also the scale at which the helicity decomposition is no longer a useful one. However, it certainly does not mean that there are ghosts, it simply means that the helicity decomposition is not a good way to analyze the system at these scales. As has already been shown in great detail \cite{deRham:2010ik}, the \stu decomposition demonstrates that for the special two parameter family of mass terms, the absence of ghosts at the scale $\Lambda_3$ and the ADM decomposition proves it to all orders (at least classically) \cite{Hassan:2011hr}. For these reasons there is no need to pursue the helicity decomposition to this scale, since it becomes redundant to previous analysis. \\

Despite the fact that the results are already known, let us sketch a few of the details to make clear the comparison between the helicity and \stu decompositions at the scale $\Lambda_3$. It is worth pointing out that at the scale $\Lambda_3$, the cubic interaction $\p^2 \breve{h} (\p A)^2$ from the EH is relevant and ought to be diagonalized,
\ba
\mathcal{L}^{(3)}_{\Lambda_3, \, \p^2 \breve{h} (\p A)^2}=-\frac{1}{4\Lambda_3^3}\breve{h}^{\alpha\beta}\Ein ^{\mu\nu}_{\alpha\beta}\p_\mu A_\gamma \p_\nu A^\gamma\,.
\ea
To diagonalize this interaction it is therefore necessary to make the field redefinition
\ba
\hat{h}\mn=\breve{h}\mn-\frac{1}{4\Lambda_3^3} \p_\mu A_\gamma \p_\nu A^\gamma\,,
\ea
and as  result we recover the fact that after this diagonalization, the helicity-2 mode at that level is nothing else but what we would have inferred from the \stu decomposition
\ba
\label{hhat}
h\mn={h}^\st \mn+\frac 16 \pi \eta\mn+\mpl (\Psi_{\mu\alpha}\Psi_\nu^{\ \alpha}-\eta\mn)\,.
\ea

{\bf We have derived here the \stu decomposition without ever invoking diffeomorphism invariance!}
To be clear we are not dealing with the \stu decomposition in its standard presentation, but rather what the \stu decomposition looks like when translated back into unitary gauge. The relationship between the above defined unitary gauge metric and the metric in \stu gauge, in which the 4 \stu fields are taken to be dynamical, is given by
\ba
g_{\mu\nu} = g^S_{\alpha \beta} \partial_{\mu} \Phi^{\alpha} \partial_{\nu} \Phi^{\beta}
\ea
With $g^S_{\alpha \beta} = \eta_{\alpha \beta}+ h^S_{\alpha \beta}$ and $\Phi^A = x^A + \frac{1}{2\mpl  m}\left( A_{\alpha} + \frac{1}{3m} \partial_{\alpha}\pi \right)$ we obtain
\ba
h_{\mu\nu} = \mpl (g_{\mu\nu} -\eta_{\mu\nu})= \mpl  \left( \Psi_{\mu \alpha} \Psi_{\mu}{}^{\alpha} - \eta_{\mu\nu} \right) + h^S_{\alpha \beta}  \partial_{\mu} \Phi^{\alpha} \partial_{\nu} \Phi^{\beta}
\ea
As long as we are working at energy scales below $\Lambda_3$ this is equivalent to
\ba
h_{\mu\nu} = \mpl (g_{\mu\nu} -\eta_{\mu\nu})= \mpl  \left( \Psi_{\mu \alpha} \Psi_{\mu}{}^{\alpha} - \eta_{\mu\nu} \right) + h^S_{\mu \nu}
\ea
Finally writing $h^S_{\mu\nu} = h_{\mu\nu}^\st  + \frac{1}{6} \pi \eta_{\mu\nu}$ gives the desired expression
\ba
h_{\mu\nu} &=& h_{\mu\nu}^\st  + \frac{1}{6} \pi \eta_{\mu\nu}+ \mpl  \left( \Psi_{\mu \alpha} \Psi_{\mu}{}^{\alpha} - \eta_{\mu\nu} \right) \nn \\
&=&  h_{\mu\nu}^\st  +\frac{1}{2m}\partial_{(\mu} A_{\nu )} + \frac{1}{3m^2} \partial_{\mu}\partial_{\nu} \pi +D^\st _{\mu\nu}\,.
\ea

In \cite{Folkerts:2011ev}, it was suggested that their apparent discrepancy between the \stu language and the helicity decomposition is related to the fact that the helicity-0 mode  in both languages is related via a nonlinear expression
\ba
\pi^\hel \eta\mn =\pi^\st  \eta\mn+\frac{1}{\Lambda_5^5}\(\p_\mu\p_\alpha \pi^\st \p_\nu\p^\alpha \pi^\st -\cdots\)\,,
\ea
where in their logic, $\pi^\hel $ identifies the helicity-0 mode in the helicity decomposition while $\pi^\st $ is the scalar that appears in the \stu language. By this point it is however obvious that this is by no mean the correct identification to be made. In reality, the mode  $\pi^\st $ is nothing else but the very same helicity-0 mode $\pi^\hel $ that appears in the helicity decomposition. Instead, the \stu and helicity decomposition are reconciled via the correct identification of the helicity-2 part of the massive spin-2 field. As we have seen, up to the scale $\Lambda_3$, the correctly diagonalized  helicity-2 mode is not $h^\hel \mn$ but instead $ h^\st \mn$ given in \eqref{hhat}. Of course there is no problem whatsoever in expressing $h^\hel \mn$ in terms of $ h^\st \mn$, their relation is perfectly well-defined and invertible.

\section{Outlook}
\label{sec:Outlook}

In this article we have considered the helicity decomposition of the two parameter allowed ghost-free interacting models of massive spin-2 fields, \ie massive gravity \cite{deRham:2010kj}. We find that, despite the apparent hierarchy of scales due to the appearance of terms in the Lagrangian at the intermediate scales $\Lambda_5$ and $\Lambda_4$, the first interactions of the helicity modes arise at the scale $\Lambda_3$. We demonstrate this result without the use of field redefinitions. We show that the equations of motion of the helicity modes have a well-defined Cauchy problem in the presence of matter, and that the symmetry that arises in the $m \rightarrow 0$ limit is sufficient to ensure that the loops from matter never reintroduce problems at the scales $\Lambda_5$ or $\Lambda_4$. This result confirms the absence of ghosts  in these decoupling limits. We show that perturbation of the spin-2 field remains well-defined below the scale $\Lambda_3$ confirming that $\Lambda_3$ is the true strong coupling scale of the helicity-0 mode. Our results are completely consistent with the ghost-free proof of these models using the ADM formalism \cite{deRham:2010kj,Hassan:2011hr} and are consistent with how the ghosts can be seen to be absent in the \stu formalism \cite{deRham:2011rn}.

\acknowledgments
We would like to thank Clare Burrage, Lavinia Heisenberg and Kurt Hinterbichler for useful discussions and insightful comments.
CdR is funded by the SNF and the work of GG was supported by NSF grant PHY-0758032. AJT would like to thank the  Universit\'e de  Gen\`eve for hospitality whilst this work was being completed.

\appendix

\section{Phase Space degrees of freedom}
\label{appendix}

In this appendix we show how the Hamiltonian of the scalar field model \eqref{toy} is perfectly well defined and propagates only four degrees of freedom in phase space.
To simplify we only focus on the time dependencies, such that the Lagrangian is of the form
\ba
\mathcal{L}=\frac 12 \(\dot \psi+\frac{2}{\Lambda^5}\ddot \phi\,  \dddot \phi\)^2+\frac 12 \dot \phi^2\,.
\ea
To understand better the degrees of freedom, it is helpful to rewrite this as a constrained system
\ba
\mathcal{L}=\frac 12 \(\dot \psi+\frac{2}{\Lambda^5}\mu \,  \dot \mu\)^2+\frac 12 \dot \phi^2 + \rho \, \left(\mu - \ddot{\phi} \right)\, .
\ea
where the auxiliary field $\rho$ enforces $\mu = \ddot{\phi}$ making transparent the equivalence with the previous system. Now on integration by parts this is equivalent to 
\ba
\mathcal{L}=\frac 12 \(\dot \psi+\frac{2}{\Lambda^5}\mu \,  \dot \mu\)^2+\frac 12 \dot \phi^2 + \rho \mu + \dot{\rho} \dot{\phi}\ea

A priori this extends the phase space to a eight dimensional one (2 times too large), signaling the presence of a ghost, however the theory at hand is (as we know) extremely special in that it propagates a constraint.
One can start by defining the conjugate momenta associated to $\phi$, $\psi$, $\rho$ and $\mu$,
\ba
P_{\psi}&=&\dot \psi+\frac{2}{\Lambda^5} \mu \dot{\mu} \\
P_\phi&=&\dot \phi + \dot{\rho}\,, \\
P_{\rho} &=& \dot{\phi} \\
P_{\mu} &=& \frac{2}{\Lambda^5} \mu \left( \dot \psi+\frac{2}{\Lambda^5} \mu \dot{\mu} \right)
\ea
Already it is clear that there is a constraint $C_1=P_{\mu} -\frac{2}{\Lambda^5} \mu P_{\psi}=0$. Taking this into account the Hamiltonian is
\be
\mathcal{H}= \left[ \frac{1}{2}P_{\psi}^2 +P_{\phi} P_{\rho}-\frac{1}{2}P_{\rho}^2-\rho \mu \right] +\lambda_1 \left(P_{\mu} -\frac{2}{\Lambda^5} \mu P_{\psi} \right)
\ee
However there is also a secondary constraint $C_2= i [C_1,H]=\rho=0$ and a tertiary constraint $C_3=i [C_2,H]=P_{\rho}-P_{\phi}=0$. Putting this together we get
\be
\mathcal{H}= \left[ \frac{1}{2}P_{\psi}^2 +\frac{1}{2} P_{\phi}^2\right] +\lambda_1 \left(P_{\mu} -\frac{2}{\Lambda^5} \mu P_{\psi} \right)+\lambda_2 ( \rho)+ \lambda_3 \left( P_{\rho}-P_{\phi} \right)
\ee
with $\lambda_1,\lambda_2,\lambda_3$ Lagrange multipliers.
Although $C_2$ and $C_3$ are second class constraints since $[C_2,C_3]=i$, $C_1$ is clearly first class and thus removes two degrees of freedom. It generates the symmetry $\psi \rightarrow \psi + 2 \mu \alpha(t)$, $\mu \rightarrow \mu-\Lambda_5^5 \alpha(t)$ for infinitessimal $\alpha(t)$. Thus there are a total of $8-2 (\text{second class})-2 \times 1 (\text{first class})=4 $ phase space degrees of freedom and hence 2 physical degrees of freedom. As usual this was manifest from the outset by performing the field redefinition $\psi = \bar{\psi}- \ddot{\phi}^2/\Lambda^5$, however since this is a non-canonical transformation it was important to check that this redefinition could be consistently performed.

Interestingly the special coupling to sources considered in Section~\ref{couplingtosources} is precisely the coupling
\be
\mathcal{L}_{\rm source} = \bar{J}_{\phi} \phi + \bar J_{\psi} \left( \psi +\frac{1}{\Lambda^5} (\p_\alpha \p_\beta \phi)^2 \right) = \bar{J}_{\phi} \phi + \bar J_{\psi} \left( \psi +\frac{1}{\Lambda^5} \mu^2 \right)
\ee 
which preserves the first class symmetry. This is yet another way to understand how the correct number of physical degrees of freedom can be maintained at the quantum level even in the presence of sources since it will be protected by this gauge symmetry.



\end{document}